\documentclass[reqno]{amsart}
\usepackage[T1]{fontenc}
\usepackage[utf8]{inputenc}
\usepackage{graphicx} 
\usepackage[svgnames,table]{xcolor}
\usepackage{amsmath}
\usepackage{amsfonts}
\usepackage{amssymb}
\usepackage{amsthm}
\usepackage{adjustbox}
\usepackage{array}
\usepackage{booktabs}
\usepackage{cite}
\usepackage{caption}
\usepackage{tikz,pgfplots}
\usepackage{float}
\pgfplotsset{compat=1.18}
\usepgfplotslibrary{fillbetween}
\usetikzlibrary{calc,arrows,arrows.meta,cd,shapes.geometric,positioning,through,decorations.markings,decorations.text}
\tikzset{>=latex}
\usepackage[margin=1in]{geometry}
\usepackage{makecell}
\usepackage{rotating}
\usepackage[countmax]{subfloat}
\usepackage{hyperref}
\newcommand{\lagg}{\mathfrak{g}}
\newcommand{\hagg}{\mathfrak{h}}

\newcommand{\agg}{\mathfrak{a}}
\newcommand{\tagg}{\mathfrak{t}}

\newcommand{\cagg}{\mathfrak{c}}
\newcommand{\bbR}{\mathbb{R}}
\newcommand{\bbC}{\mathbb{C}}
\newcommand{\bbZ}{\mathbb{Z}}
\newcommand{\bbN}{\mathbb{N}}
\newcommand{\sg}{\mathsf{g}}
\newcommand{\sk}{\mathsf{k}}
\newcommand{\sh}{\mathsf{h}}
\newcommand{\sX}{\mathsf{X}}
\newcommand{\sZ}{\mathsf{Z}}
\newcommand{\sM}{\mathsf{M}}
\newcommand{\sY}{\mathsf{Y}}
\newcommand{\st}{\mathsf{t}}
\newcommand{\maH}{\mathcal{H}}
\newcommand{\bv}{\boldsymbol{v}}
\newcommand{\bx}{\boldsymbol{x}}
\newcommand{\bp}{\boldsymbol{p}}
\newcommand{\be}{\begin{equation}}
\newcommand{\ee}{\end{equation}}
\DeclareMathOperator{\Spin}{Spin}

\DeclareMathOperator{\ISO}{ISO}
\DeclareMathOperator{\Aff}{Aff}

\DeclareMathOperator{\U}{U}
\DeclareMathOperator{\SU}{SU}
\DeclareMathOperator{\PU}{PU}
\DeclareMathOperator{\SO}{SO}
\DeclareMathOperator{\SL}{SL}
\DeclareMathOperator{\Euc}{Euc}
\DeclareMathOperator{\Lie}{Lie}
\DeclareMathOperator{\GL}{GL}

\DeclareMathOperator{\N}{N}
\DeclareMathOperator{\sgn}{sgn}
\DeclareMathOperator{\Ad}{Ad}
\DeclareMathOperator{\ad}{ad}
\newtheorem{theorem}{Theorem}[section]

\newsavebox{\pmatrixbox}
\newenvironment{colourpmatrix}
  {\begin{lrbox}{\pmatrixbox}
   \mathsurround=0pt
   $\displaystyle
   \begin{pmatrix}}
  {\end{pmatrix}$%
   \end{lrbox}%
   \usebox{\pmatrixbox}%
   \kern-\wd\pmatrixbox
   \makebox[0pt][l]{$\left(\vphantom{\usebox{\pmatrixbox}}\right.$}%
   \kern\wd\pmatrixbox
}
\title{Elementary Classical and Quantum Lifshitz Systems}
\author{Jarah Fluxman}
\address{Maxwell Institute and School of Mathematics, The University of Edinburgh, James Clerk Maxwell Building, Peter Guthrie Tait Road, Edinburgh EH8 3FD, Scotland, United Kingdom}
\email{J.C.Fluxman@sms.ed.ac.uk}

\numberwithin{equation}{section}
\begin{document}

\begin{abstract}
    We classify the elementary classical and quantum Lifshitz systems. Lifshitz systems are systems where space and time scale anisotropically. That is, there is a constant $z$ such that under scaling by a factor of $\lambda$,
    \begin{equation*}
    \boldsymbol{x}\rightarrow \lambda \boldsymbol{x} \text{ and } t\rightarrow \lambda^{z}t
    \end{equation*}
    There are seven Lie groups, called the Lifshitz groups, which characterise all Lifshitz symmetries.
    Elementary classical Lifshitz systems are the symplectic manifolds with a transitive Lifshitz action, which turn out to be coadjoint orbits of the Lifshitz groups and their one-dimensional central extensions up to covering. Elementary quantum Lifshitz systems are the projective unitary irreducible representations of the Lifshitz groups. 
\end{abstract}

\maketitle
\tableofcontents
\section{Introduction}
In recent years there has been research into physical systems that exhibit Lifshitz symmetry, namely anisotropic scaling of spatial and time dimensions, or of spatial dimensions in different directions. The two principal areas in which this has been investigated are holography \cite{MR3529553} \cite{Baggio:2011ha} \cite{MR2551709}, and condensed matter \cite{Inkof:2019gmh}. There have been attempts to link the two and build holographic condensed matter theories \cite{Zaanen:2015oix}. The basic principle underlying these areas of research is an anisotropic scaling of position and time:
\begin{equation}
    \boldsymbol{x}\rightarrow \lambda \boldsymbol{x} \text{ and } t\rightarrow \lambda^{z}t
\end{equation}
where $z$ is a real parameter. When $z=1$ the scaling is isotropic and spacetime symmetry is restored. The Lie groups underlying this type of scaling are called the Lifshitz Lie groups. Lifshitz Lie groups are generated by spatial rotations, translations, a time translation and a dilation, the last of which provides the anisotropic scaling. 

As an example, consider the spacetime metric from \cite{MR3529553}:
\begin{equation}
    ds^2=\frac{dr^2}{r^2}+\frac{dx^i dx_i}{r^2}-\frac{dt^2}{r^{2z}}
\end{equation}
This metric is invariant under Lifshitz scaling $r\rightarrow \lambda r$, $x^i\rightarrow \lambda x^i$, $t\rightarrow \lambda^z t$.

The prominence of Lifshitz systems in research has brought to light the need to classify all classical and quantum dynamical systems that have Lifshitz symmetry. A classical system exhibiting Lifshitz symmetry is mathematically defined in the sense of Souriau \cite{MR1461545} as a symplectic manifold on which a Lifshitz group acts transitively and symplectically.  The information encoded in such a symplectic manifold can be used to generate curves in a Lifshitz system, following a Lifshitz geometry such as the one above. It has been shown in \cite{MR1461545}\cite{MR412321}\cite{MR294568}\cite{Beckett:2022wvo} that every simply-connected symplectic manifold on which a connected Lie group acts transitively and symplectically covers a coadjoint orbit of either that group or one of its one-dimensional central extensions. The sympectic form is the pullback of the Kirillov-Kostant-Souriau symplectic form on the coadjoint orbit and the symplectic group action corresponds to the coadjoint action. A quantum Lifshitz system is defined à la Wigner \cite{MR1503456} as a projective unitary representation of a Lifshitz group. Projective representations of a Lie group lift to true representations of a one-dimensional central extension of that group. UIRs of Lifshitz groups give quantum fields on Lifshitz spacetime under certain transformations.

This article's purpose is to completely classify the elementary classical and quantum Lifshitz systems in 2 and 3 dimensions. This amounts to classifying the coadjoint orbits and unitary irreducible representations of the Lifshitz groups and their one-dimensional central extensions. There is a rough correspondence between coadjoint orbits and UIRs via geometric quantisation but this isn't always the case. The correspondence holds for some classes of Lie group such as nilpotent groups, compact groups and those of the form $K\ltimes T$, where $T$ is abelian, but not always for general Lie groups \cite{MR1486137}\cite{MR2069175}\cite{MR387499}\cite{Gravog}. For example, the complementary series UIRs for the group $\SL(2,\bbC)$ do not arise from quantising a coadjoint orbit. Not every Lifshitz group is of a form where the correspondence between coadjoint orbits and UIRs is known to hold. Nevertheless, in so far as it is possible, we would like to build a ``dictionary'' between Lifshitz coadjoint orbits and Lifshitz UIRs. 

We begin in section 2 by reviewing the Lifshitz algebras and their central extensions. There are seven Lifshitz algebras up to isomorphism and these are listed in table \ref{tab1}. They were obtained by extending Aristotelian algebras by a grading element in \cite{MR4577534}. The central extensions of the Lifshitz algebras are listed in table \ref{tab2}. Many of the Lifshitz algebras and their central extensions decompose as sums of simpler ``building-block'' algebras. In such cases, the coadjoint orbits are the products of the orbits of the corresponding building-block groups and the UIRs are the tensor products of the UIRs of the building-block groups. There are also the ``indecomposable'' algebras, which do not break down into sums of simpler algebras. These are the algebras labelled $\agg_3^z$, $\agg_7$, $\cagg_1^1$ and $\cagg_3$. 

In section 3 we discuss coadjoint orbits of a Lie group $G$ as classical systems with $G-$symmetry and unitary irreducible representations of $G$ as quantum systems with $G-$symmetry. We also motivate why it is necessary not just to consider coadjoint orbits and UIRs of $G$, but also of the one-dimensional central extensions of $G$. Finally we discuss the method of Mackey, which is the key tool we will use to classify the UIRs of the Lifshitz groups and their central extensions. Mackey's method allows us to find all the UIRs of a semidirect product group $G=K\ltimes T$ from those of $T$ and certain subgroups of $K$. Mackey's method is fairly simple when $T$ is abelian and all the UIRs of $T$ are one-dimensional but becomes more complex when $T$ is not abelian and has higher dimensional UIRs. In such a case we require a unitary intertwiner $S$ between equivalent representations of $T$, which is sometimes difficult to determine. Most groups discussed in this paper have abelian $T$ but a few do not. These are the oscillator group $\mathcal{D}$ and the groups denoted $A_7$ and $C_1^1$. 

Section 4 is concerned with the building-block groups. The UIRs and coadjoint orbits of these groups are known and can be found in the literature. Nevertheless, we review them all under one umbrella to establish notational consistency. The principal results are given in subsection \ref{results1}. Table \ref{tab3} lists all the coadjoint orbits of the building-block groups while table \ref{tab4a} and table \ref{tab4b} list their UIRs. 

Section 5 contains new research, namely the classification of the coadjoint orbits and UIRs of the indecomposable groups, those groups which don't break up into simpler building-block groups. The principal results and their interpretations are given in subsection \ref{results2}. Table \ref{tab5} lists all the coadjoint orbits of the indecomposable groups while table \ref{tab6a} and table \ref{tab6b} list their UIRs. With the exception of the complementary series for the bulding-block groups $\SL(2,\bbC)$ and $\SL(2,\bbR)$, the coadjoint orbits and the UIRs of all building-block and indecomposable groups are in exact correspondence.

\section{The Lifshitz Algebras and their Central Extensions}

A $d-$dimensional Lifshitz algebra is a real Lie algebra spanned by the generators $\{L_{ab},P_c,H,D\}$, with $a,b,c \in \{1,2,..,d\}$. The generators $L_{ab}$ span an $\mathfrak{so}(d)-$subalgebra, which acts on $P_a$ in the vector representation and on $H$ and $D$ in the scalar representation. The brackets involving these ``rotation'' generators are

\begin{equation}
\begin{split}
   [L_{ab},L_{cd}] &= \delta_{ad}L_{bc}-\delta_{ac}L_{bd}+\delta_{bc}L_{ad}-\delta_{bd}L_{ac}\\
   [L_{ab},P_c]&= \delta_{bc}P_a -\delta_{ac}P_b\\
   [L_{ab},H]&= 0\\
   [L_{ab},D]&= 0
\end{split}
\end{equation}
The Lifshitz algebras are classified up to isomorphism by considering all bracket relations on $\{P_a,H,D\}$ which are compatible with the above in the sense of the Jacobi identity. In \cite{MR4577534} it was shown that every Lifshitz algebra can be otained by centrally extending an Aristotelian algebra by a grading element. The Lifshitz algebras were thus classified up to isomorphism:

\begin{table}[h!]
  \centering
  \caption{Lifshitz algebras up to isomorphism}
  \label{tab1}
  \rowcolors{2}{blue!10}{white}
\begin{tabular}{w{c}{1cm}|c|c|c|c|w{c}{1cm}|w{c}{3cm}} 
 Name & $d$&$[D,H]$ & $[D,P_a]$&$[H,P_a]$& $[P_a,P_b]$& Comments\\
 \toprule
 $\mathfrak{a}_1 $ &$\geq 2$&$0$&$0$& $0$ & $0$& $\bbR^2\oplus\mathfrak{iso}(d)$\\ 
 $\mathfrak{a}_2$  &$\geq 2$&$H$&$0$ &$0$ & $0$& $\mathfrak{aff}(1)\oplus\mathfrak{iso}(d)$\\
 $\mathfrak{a}_3^z$  &$\geq 2$&$zH$&$P_a$& $0$ & $0$ & $z \in \bbR$\\ 
 \makecell{$\agg^+_4$ \\ $\agg^-_4$} &$\geq 2$ &$0$&$0$& $0$ &\makecell{$L_{ab}$ \\ $-L_{ab}$} & \makecell{$\bbR^2 \oplus \mathfrak{so}(d+1)$ \\ $\bbR^2 \oplus \mathfrak{so}(d,1)$}\\ 
 \makecell{$\agg^+_5$ \\ $\agg^-_5$}&$\geq 2$ &$H$&$0$& $0$ & \makecell{$L_{ab}$ \\ $-L_{ab}$}&\makecell{$\mathfrak{aff}(1) \oplus \mathfrak{so}(d+1)$ \\ $\mathfrak{aff}(1) \oplus \mathfrak{so}(d,1)$}\\
 $\mathfrak{a}_6$&$2$&$0$&$0$&$0$&$\epsilon_{ab}H$&$\bbR\oplus\mathfrak{d}$\\
 $\mathfrak{a}_7$&$2$&$2H$&$P_a$&$0$&$\epsilon_{ab}H$& $(\mathfrak{so}(2)\oplus\bbR)\ltimes \mathfrak{n}$\\
 \bottomrule
\end{tabular}
\caption*{Some Lifshitz algebras break up into direct sums of simpler building-block algebras: $\mathfrak{iso}(d)$ is the $d-$dimensional euclidean algebra; $\mathfrak{aff}(1)$ denotes the two-dimensional non-abelian algebra (or the algebra of the one-dimensional affine group); $\mathfrak{d}$ is the ``diamond'' or ``oscillator'' algebra; $\mathfrak{so}(d+1)$ and $\mathfrak{so}(d,1)$ are respectively the special orthogonal and lorentzian algebras; and $\mathfrak{n}$ is the Heisenberg algebra.}
\end{table}
The first three algebras $\agg_1$, $\agg_2$ and $\agg_3^z$ should be seen as part of the same family of algebras characterised by the adjoint action of $D$ on $P_a$ and $H$. The algebra where this action is trivial is $\agg_1$. The parameter $z$ can be thought of as a homogeneous coordinate in the projective space of scalings of $P_a$ and $H$ under the $D-$action, with $\agg_2$ covering the case where $z\rightarrow \infty$.

The homogeneous spaces of the simply-connected Lie groups corresponding to the Lifshitz algebras were classified by their Klein pairs in \cite{MR4577534}. Of particular interest are the so-called Lifshitz-Weyl spacetimes, which are characterised by the Klein pair $(\lagg,\mathfrak{so}(d)\oplus \bbR D)$, where $\lagg$ is a general Lifshitz algebra. These spaces are called spacetimes because they can be locally coordinatised by the coset representative $\exp(x^a P_a+tH)$, where $x^a$ and $t$ are viewed as spatial and temporal coordinates respectively. The generator $D$ is then a dilation, which scales space and time anisotropically. 

Many of the Lifshitz algebras break down into direct sums of simpler algebras. In the algebra $\agg_1$, the generators $D$ and $H$ generate an abelian subalgebra $\bbR^2$. This algebra commutes with the other generators, $\{L_{ab},P_a\}$, which generate a \textbf{euclidean} $\mathfrak{iso}(d)$ subalgebra. Hence $\agg_1\cong\bbR^2\oplus \mathfrak{iso}(d)$ as a Lie algebra. In the algebra $\agg_2$, the generators $D$ and $H$ also generate a subalgebra, although this time it is the unique \textbf{non-abelian} or \textbf{affine} algebra $\mathfrak{aff}(1)$. So $\agg_2\cong\mathfrak{aff}(1)\oplus \mathfrak{iso}(d)$. The generators $D$ and $H$ also generate subalgebras of $\agg_4^\pm$ and $\agg_5^\pm$, although this time $\{L_{ab},P_a\}$ do not form a euclidean subalgebra because the $P_a$ no longer commute. Instead, the Lie bracket between the $P_a$ gives rotation generators. Depending on whether the sign in the commutation relation is $+$ or $-$, the algebra generated by $\{L_{ab},P_a\}$ is either the \textbf{special orthogonal} algebra $\mathfrak{so}(d+1)$ or the \textbf{lorentzian} algebra $\mathfrak{so}(d,1)$. The Lifshitz-Weyl spacetimes generated by $\agg_4^\pm$ and $\agg_5^\pm$ are therefore spherical or hyperbolic. The algebra $\agg_6$ only exists in dimension $d=2$. It contains a subalgebra $\mathfrak{d}$, generated by $\{L,P_1,P_2,H\}$, where $L$ is the one-dimensional generator of the $\mathfrak{so}(2)-$subalgebra. $\mathfrak{d}$ is the \textbf{diamond} or \textbf{oscillator} algebra. The dilation generator $D$ commutes with this $\mathfrak{d}$, so we see that $\agg_6\cong\bbR\oplus\mathfrak{d}$. There are two algebras which do not break into direct sums of simpler algebras. These are $\agg_3^z$, which is more properly a family of algebras parametrised by the real number $z$, and $\agg_7$.

In order to fully classify the elementary classical and quantum Lifshitz systems, we have to consider not only the Lifshitz algebras but their one-dimensional central extensions as well. A one-dimensional central extension of a Lie
algebra extends the algebra by one additional generator, which commutes
with the rest. That is, given an algebra $\lagg$, we construct an
algebra $\hat{\lagg}$ of one additional dimension such that there is a
short exact sequence of Lie algebra homomorphisms:

\begin{equation}
  \begin{tikzcd}
    0 \arrow[r] & \bbR \arrow[r, "i"] & \hat{\lagg} \arrow[r] & \lagg \arrow[r] & 0
 \end{tikzcd}
\end{equation}
such that $i(\bbR)$ is central in $\hat{\lagg}$. Two central extensions $\hat{\lagg}_1$ and $\hat{\lagg}_2$ of an algebra $\lagg$ are said to be \textbf{equivalent} if there is an algebra isomorphism $\psi:\hat{\lagg}_1\rightarrow\hat{\lagg}_2$ such that the following diagram commutes:

\begin{equation}
  \begin{tikzcd}
    0 \arrow[r] & \bbR \arrow[d, equal] \arrow[r, "i_1"] & \hat{\lagg}_1 \arrow[d, "\psi"] \arrow[r] & \lagg \arrow[d, equal]\arrow[r] & 0\\
    0 \arrow[r] & \bbR \arrow[r,"i_2"] & \hat{\lagg}_2\arrow[r]&\lagg \arrow[r] &0
 \end{tikzcd}
\end{equation}
It is always possible to trivially extend a Lie algebra simply by letting $\hat{\lagg}=\lagg\oplus\bbR$. However, certain algebras also admit non-trivial one-dimensional central extensions, which are not equivalent to this construction. The one-dimensional central extensions of a Lie algebra are classsified up to equivalence by its second Chevalley-Eilenberg cohomology with values in the trivial one-dimensional representation.

The one-dimensional central extensions of the Lifshitz algebras were calculated in \cite{MR4577534}. There is a subtley in that not every central extension of a Lie group's Lie algebra integrates to a central extension of the group itself, although this is true when the group is simply connected \cite{MR1424633}. Thus in dimension $d=2$, there are central extensions of the Lifshitz algebras which are not central extensions of the Lifshitz groups. The following table lists all non-trivial central extensions of the Lifshitz algebras which also integrate to central extensions of the groups. 

\begin{table}[hbt!]
  \centering
  \caption{Central extensions which integrate to extensions of their groups}
  \label{tab2}
  \rowcolors{2}{blue!10}{white}
  \begin{tabular}{w{c}{1cm}|w{c}{3cm}|c|c|c} 
    Name & Central Extension of& $d$& Brackets involving $Z$&Comments\\
    \midrule
    $\cagg_1$&$\mathfrak{a}_1 $ &$\geq 2$& $[D,H] = Z$&$\mathfrak{iso}(d)\oplus \mathfrak{n}$\\ 
    $\cagg_1^0$&$\mathfrak{a}_1$ &$2$& $[P_a,P_b]= \epsilon_{ab}Z$&$\bbR^2\oplus\mathfrak{d}$\\
    $\cagg_1^{1}$&$\mathfrak{a}_1$& $2$ & $[D,H]= Z$, $[P_a,P_b]=\epsilon_{ab}Z$ &\\
    $\cagg_2 $&$\mathfrak{a}_2$  &$2$& $[P_a,P_b]=\epsilon_{ab}Z$&$\mathfrak{aff}(1)\oplus\mathfrak{d}$\\
    $\cagg_3 $&$\mathfrak{a}_3^{z = 0}$  &$\geq 2$& $[D,H]=Z$&\\ 
    \makecell{$\cagg^+_4$ \\ $\cagg^-_4$}&\makecell{$\agg^+_4$ \\ $\agg^-_4$}&$\geq 2$ &$[D,H]=Z$&  \makecell{$\mathfrak{n} \oplus \mathfrak{so}(d+1)$ \\ $\mathfrak{n} \oplus \mathfrak{so}(d,1)$}
    \\ 
    \bottomrule
  \end{tabular}
  \caption*{Many of the central extensions also break down as direct sums of simpler algebras.}
\end{table}

In the algebra $\cagg_1$, the generators $D,H$ and $Z$ span a \textbf{Heisenberg} algebra $\mathfrak{n}$, which commutes with the $\mathfrak{iso}(d)$ subalgebra spanned by the rotation and translation generators. Thus $\cagg_1\cong\mathfrak{n}\oplus\mathfrak{iso}(d)$. In the algebra $\cagg_1^0$, which only exists in 2 dimensions, the generators $\{L,P_1,P_2,Z\}$ generate an oscillator algebra, while $D$ and $H$ again span an abelian algebra $\bbR^2$. Thus $\cagg_1^0\cong\bbR^2\oplus\mathfrak{d}$. The situation is much the same for the algebra $\cagg_2$, except this time $D$ and $H$ span a non-abelian algebra. Finally, in $\cagg_4^\pm$, $D,H$ and $Z$ span a Heisenberg algebra, while $\{L_{ab},P_a\}$ again span a special orthogonal or lorentzian algebra, so that $\cagg_4^+ \cong \mathfrak{n}\oplus \mathfrak{so}(d+1)$ and $\cagg_4^- \cong \mathfrak{n}\oplus \mathfrak{so}(d,1)$. The algebras $\cagg_1^1$ and $\cagg_3$ do not break down into direct sums of building-block algebras.

Now that the Lifshitz algebras and their central extensions have been examined, the question remains as to the Lie groups they generate. In this article we restrict ourselves to the groups of low dimension ($d=2$ and $d=3$) to maintain simplicity of calculation. Each algebra is of the form $(\mathfrak{so}(d)\oplus \bbR D)\ltimes \mathfrak{t}$ for some algebra $\mathfrak{t}$. We define the Lifshitz Lie group (or central extension) to be $(\Spin(d)\times \bbR)\ltimes T$, where $T$ is the unique simply-connected Lie group with algebra $\mathfrak{t}$. In particular this means that the groups are not simply-connected for $d=2$ because $\Spin(2)\cong \U(1)$ is not simply connected.

\section{Coadjoint Orbits and UIRs as Classical and Quantum Lifshitz Systems}
\subsection{Coadjoint Orbits as Symplectic $G-$manifolds}
A classical $G-$system, where $G$ is a Lie group, is defined as a symplectic manifold admitting a transitive $G$-action, which is also a symplectomorphism \cite{MR1461545}. The coadjoint orbits of $G$ are themselves symplectic manifolds on which $G$ acts transively and symplectomorphically. Each coadjoint orbit is endowed with a $G-$invariant closed two-form $\omega_{KKS}$ called the \textbf{Kirillov-Kostant-Souriau} symplectic form. The construction of $\omega_{KKS}$ is given below. Since the symplectic forms are not actually calculated for each coadjoint orbit in this paper, a reader wishing simply to understand the main results need only consider equations \ref{ad} and \ref{coad}. Coadjoint orbits are not simply examples of classical $G-$systems, they classify them completely. It is a result that every symplectic $G-$manifold is in fact the universal cover of a coadjoint orbit \cite{MR1461545}\cite{MR412321}\cite{MR294568}\cite{Beckett:2022wvo}. This allows us to understand classical $G-$systems simply by classifying coadjoint orbits. Much of this subsection is taken from \cite{Beckett:2022wvo} and \cite{primer} and the interested reader is referred to those two papers for more details.

Recall that any Lie group $G$ has an action on its own Lie algebra, $\lagg$, called the \textbf{adjoint action}. Briefly, let $\sg \in G$ and $\sX\in \lagg$.  Then the adjoint action is defined as
\begin{equation}\label{ad}
    \Ad_\sg \sX := \frac{d}{ds}g \exp(s\sX) g^{-1}|_{s=0}
\end{equation}
The \textbf{coadjoint action} is an action of $G$ on its coalgebra, $\lagg^*$, which is dual to the adjoint action. Let $\sg \in G$, $\sX\in \lagg$ and $\sM \in \lagg^*$. Let $\langle,\rangle$ be the pairing between $\lagg^*$ and $\lagg$. Then the coadjoint action $\Ad^*$ is defined by
\begin{equation}\label{coad}
    \langle \Ad^*_\sg \sM, \sX\rangle := \langle \sM,\Ad_{\sg^{-1}}\sX\rangle
\end{equation}
The coalgebra $\lagg^*$ is thus partitioned into $G-$orbits, called the \textbf{coadjoint orbits} of $G$. Since the coadjoint action is smooth the coadjoint orbits are submanifolds of $\lagg^*$. The adjoint action linearises to an action of $\lagg$ on itself. Let $\sX,\sY\in \lagg$. Then
\begin{equation}
    \ad_\sX \sY := \frac{d}{ds}\Ad_{\exp{s\sX}}\sY |_{s=0}=[\sX,\sY]
\end{equation}
There is also a linearisation of the coadjoint action:
\begin{equation}
    \ad^*_{\sX}\sM=-\sM\circ \ad_\sX
\end{equation}
Let $H$ be the stabiliser of $\sM \in \lagg^*$ under the coadjoint action and $\mathcal{O}_\sM$ be its coadjoint orbit.
\begin{equation}
\begin{split}
    H &:= \{\sg\in G|\Ad^*_\sg \sM =\sM\}\\
    \mathcal{O}_\sM &:=\{\Ad^*_\sg \sM|\sg \in G\}
\end{split}
\end{equation}
Then $\mathcal{O}_\sM$ is a homogeneous space of $G$:
\begin{equation}
    \mathcal{O}_\sM =G/H
\end{equation}
Let $\hagg :=\Lie(H)$. For all $\sX\in \hagg$,
\begin{equation}
    \ad^*_\sX \sM =0
\end{equation}
The map $\alpha_\sM: G \rightarrow \mathcal{O}_\sM$, which sends $\sg\in G$ to $\Ad^*_{\sg}\sM \in \mathcal{O}_\sM$ linearises to a map $d\alpha_\sM: \lagg \rightarrow T_\sM \mathcal{O}_\sM$, which goes as 
\begin{equation}
    d\alpha_M(\sX):=\frac{d}{ds}\Ad^*_{\exp{s \sX}}\sM |_{s=0}
\end{equation}
On the other hand, the map $\alpha_\sg: \mathcal{O}_\sM\rightarrow \mathcal{O}_\sM$, which sends $p\in \mathcal{O}_\sM$ to $\Ad^*_\sg p$ for a specific $\sg$, linearises to $d\alpha_\sg: T\mathcal{O}_\sM \rightarrow T \mathcal{O}_\sM$. Since $H$ stabilises $\sM$, this linearisation turns $T_\sM \mathcal{O}_\sM$ into a representation of $H$, called the \textbf{linear isotropy representation}. For $\bv \in \sM$ and $\sh\in H$,
\begin{equation}
    \sh\cdot \bv := d\alpha_{\sh}(\bv)
\end{equation}
The map $d\alpha_\sM$ intertwines between the adjoint action of $H$ on $\lagg$ and the linear isotropy representation:
\begin{equation}
    \begin{split}
        d\alpha_\sM (\Ad_\sh \sX)&=\frac{d}{ds}\Ad^*_{\exp s \Ad_\sh \sX}\sM|_{s=0}\\
        &=\frac{d}{ds}\Ad^*_{\sh \exp{s\sX}\sh^{-1}}\sM|_{s=0}\\
        &=\frac{d}{ds}\Ad^*_\sh \Ad^*_{\exp{s\sM}}\Ad^*_{\sh^{-1}}\sM|_{s=0}\\
        &=\frac{d}{ds}\Ad^*_\sh \Ad^*_{\exp{s\sM}}\sM|_{s=0}\\
        &=\sh \cdot d\alpha_\sM(\sX)
    \end{split}
\end{equation}
The kernel of the map $d\alpha_\sM$ is precisely $\hagg$, which shows that $T_\sM \mathcal{O}_\sM \cong \lagg/\hagg$ as $H-$modules. 

We state a known result homogenous spaces of a Lie group $G$. Let $M$ be a homogeneous space of $G$ and $p\in M$. Let $H$ be the stabiliser of $p$ in $G$. Then $M\cong G/H$. Furthermore, any tensor at $p$ (that is element of $T_p M$, its dual or any of their tensor products), which is invariant under the linear isotropy action of $H$, corresponds to a $G-$invariant tensor field on $M$. Briefly, let $\omega$ be a $G-$invariant tensor field on $M$. Then by definition $\omega_{p}$ is $H-$invariant. Conversely, let $\omega_p$ be an $H-$invariant tensor at $p$. Then define $\omega$, a $G-$invariant tensor field on $M$ by $\omega_{\sg\cdot p}:=d\alpha_{\sg}(\omega_p)$.

We thus construct the symplectic form $\omega_{KKS}$ on $\mathcal{O}_{\sM}$ by considering an element $\omega_\sM$ of $\Lambda^2 T^*_\sM \mathcal{O}_\sM$ which, by the equivalence above we can consider to be a two-form on $\lagg/\hagg$. Let $\rho:\lagg\rightarrow \lagg/\hagg$ be the canonical projection. Let $\tilde{\sX}$ and $\tilde{\sY}$ be elements of $\lagg/\hagg$. Choose $\sX$ and $\sY$ in $\lagg$ such that $\tilde{\sX}=\rho(\sX)$ and $\tilde{\sY}=\rho(\sY)$. Then 
\begin{equation}
    \omega_M(\tilde{\sX},\tilde{\sY}):=\sM([\sX,\sY])
\end{equation}
This two-form is well-defined as it is independent of the choice of $\sX$ and $\sY$ in $\lagg$. Indeed let $\sX'$ and $\sY'$ be such that $\rho(\sX')=\tilde{\sX}$ and $\rho(\sY')=\tilde{\sY}$. Then $\sY'-\sY, \sX'-\sX \in \hagg$. We have 
\begin{equation}
\begin{split}
      \sM([\sX',\sY'])&=\sM([\sX+(\sX'-\sX),\sY+(\sY'-\sY)]\\
      &=\sM([\sX,\sY])+\sM([\sX,\sY'-\sY])+\sM([\sX'-\sX,\sY])+\sM([\sX'-\sX,\sY'-\sY])\\
      &=\sM([\sX,\sY])+\ad^*_{\sY'-\sY}\sM(\sX)-\ad^*_{\sX'-\sX}\sM(\sY)-\ad^*_{\sX'-\sX}\sM(\sY'-\sY)\\
      &=\sM([\sX,\sY])
\end{split}
\end{equation}
We show that $\omega_\sM$ is $H-$invariant. Let $\sh \in H$. Then
\begin{equation}
    \begin{split}
        [\sh \cdot \omega_\sM](\tilde{\sX},\tilde{\sY})&=\sM([\Ad_{\sh^{-1}} \sX,\Ad_{\sh^{-1}} \sY])\\
        &=\sM(\Ad_{\sh^{-1}}[\sX,\sY])\\
        &=\Ad^*_{\sh}\sM ([\sX,\sY])\\
        &=\omega_\sM(\tilde{\sX},\tilde{\sY})
    \end{split}
\end{equation}
The two-form $\omega_\sM$ thus gives rise to a $G-$invariant two-form on $\mathcal{O}_\sM$, which we call $\omega_{KKS}$, defined as follows. Let $\boldsymbol{x}$ and $\boldsymbol{y}$ be vectors based at $\alpha_\sg (\sM) \in \mathcal{O}_\sM$. Then 
\begin{equation}
    \omega_{KKS}(\boldsymbol{x},\boldsymbol{y}):=\omega_\sM(d\alpha_{\sg^{-1}}(\boldsymbol{x}),d\alpha_{\sg^{-1}}(\boldsymbol{y}))
\end{equation}
We note that $\omega_{KKS}$ is closed. Let $\boldsymbol{x},\boldsymbol{y}$ and $\boldsymbol{z}$ be vector fields on $\mathcal{O}_\sM$. Since the action of $G$ on $\mathcal{O}_\sM$ is transitive we can assume without loss of generality that they are generated by $\sX, \sY, \sZ \in \lagg$ respectively. Then
\begin{equation}
    d\omega_{KKS}(\boldsymbol{x},\boldsymbol{y},\boldsymbol{z})=\sum_{\text{cyclic}}\boldsymbol{x}(\omega_{KKS}(\boldsymbol{y},\boldsymbol{z}))-\sum_{\text{cyclic}}\omega_{KKS}([\boldsymbol{x},\boldsymbol{y}],\boldsymbol{z})
\end{equation}
The two-form $\omega_{KKS}$ is $G-$invariant so the first term vanishes. For the second term,
\begin{equation}
    \begin{split}
        \sum_{\text{cyclic}}\omega_{KKS}([\boldsymbol{x},\boldsymbol{y}],\boldsymbol{z})
        &=\sum_{\text{cyclic}}\omega_\sM(d\alpha_{\sg^{-1}}[\boldsymbol{x},\boldsymbol{y}],d\alpha_{\sg^{-1}}\boldsymbol{z})\\
        &=\sum_{\text{cyclic}}\sM([[\sX,\sY],\sZ])
    \end{split}
\end{equation}
which vanishes by the Jacobi identity. To prove the non-degeneracy of $\omega_{KKS}$ we simply prove the non-degeneracy of $\omega_\sM$. Let $\tilde{\sX}\in \lagg/\hagg$ be such that $\omega_\sM(\tilde{\sX},\tilde{\sY})=0$ for all $\tilde{\sY}\in \lagg/\hagg$. Then there exists $\sX\in \lagg$ such that $\rho(\sX)=\tilde{\sX}$ and $\sM([\sX,\sY])=0$ for all $\sY \in \lagg$. In other words, $\ad^*_{\sX}\sM =0$, which means that $\sX\in \hagg$. But this means that $\tilde{\sX}=\rho(\sX)=0$, so $\omega_\sM$ is non-degenerate.

The form $\omega_{KKS}$ is thus a closed, non-degenerate $G-$invariant two-form on $\mathcal{O}_\sM$, which shows that the coadjoint orbit $\mathcal{O}_\sM$ is a symplectic $G-$manifold.

Furthermore, it is a result of \cite{MR1461545}\cite{MR412321}\cite{MR294568}, restated in \cite{Beckett:2022wvo}, that every simply-connected symplectic $G-$manifold is the universal cover of a coadjoint orbit of either $G$ or one of its one-dimensional central extensions. Let $M$ be a simply-connected symplectic $G-$manifold with symplectic form $\omega$. Every $\sX \in \lagg$ gives rise to a vector field $\boldsymbol{x}$ on $M$ defined at each $p\in M$ as
\begin{equation}
    \boldsymbol{x}_p :=\frac{d}{ds}Ad^*_{\exp{s\sX}}p|_{s=0}
\end{equation}
Since $G$ acts symplectomorphically, $\mathcal{L}_{\boldsymbol{x}}\omega=0$, and since $\omega$ is symplectic, $d\omega=0$. It thus follows from Cartan's magic formula that $i_{\boldsymbol{x}}\omega$ is closed. $M$ is simply-connected so it follows that $i_{\boldsymbol{x}}\omega=d\phi_\sX$ for some function $\phi_\sX$ on $M$. There is thus a map $\phi: \lagg \rightarrow C^\infty(M)$, called the \textbf{comoment map}, defined up to a constant as
\begin{equation}
    \phi(\sX)=\phi_\sX
\end{equation}
Dual to the comoment map we have the \textbf{moment map}, $\mu:M\rightarrow \lagg^*$. This is defined as
\begin{equation}
    \langle \mu(p),\sX\rangle:=\phi_\sX(p)
\end{equation}
If the moment map intertwines with the $G-$action, that is if $\mu(\sg\cdot p)=\Ad^*_\sg \mu(p)$, then the moment map is a covering map from $M$ to $\mathcal{O}_{\mu(p)}$. $M$ is thus a cover of a coadjoint orbit of $G$. However, this is not always the case. It happens that the difference $\theta(\sg,p)=\mu(\sg\cdot p)-\Ad^*_\sg \mu(p)$ is independent of $p$ and determines a cocycle in the group cohomology of $G$. Non-trivial group cocycles correspond to central extensions of $G$ and so we get the following result:
\begin{theorem}
Let $(M,\omega)$ be a simply-connected symplectic manifold admitting a transitive action of a connected Lie group $G$ via symplectomorphisms. Then $(M,\omega)$ is the universal cover of a coadjoint orbit of $G$ or one of its one-dimensional central extensions
\end{theorem}
To understand the classical Lifshitz systems it is therefore sufficient to understand the coadjoint orbits of the Lifshitz groups and their one-dimensional central extensions.

\subsection{UIRs as quantum $G-$systems}
Much of this subsection comes directly from \cite{Simms} and the reader is referred there for a more in-depth discussion. The space of quantum mechanics is often said to be a complex Hilbert space, $\mathcal{H}$. This is not quite true, as all quantum states must be normalised to $1$. This means that the arena of quantum mechanics is the projective Hilbert space $P\maH:=\maH/\bbC^*$. 

We thus arrive at the definition of an \textbf{quantum system of a Lie group $G$}: a unitary projective representation of $G$. This is a group homomorphism $\rho:G\rightarrow \PU(\mathcal{H})$ for some Hilbert space $\mathcal{H}$. $\PU(\mathcal{H})$ is called the \textbf{projective unitary group of $\mathcal{H}$} and is defined as $\PU(\mathcal{H}):=\U(\mathcal{H})/\U(1)$.This corresponds to a map $\rho':G\rightarrow \U(\maH)$ which is a homomorphism up to a phase factor:
\begin{equation}
    \rho'(g_1g_2):=\rho'(g_1)\rho'(g_2)\exp{i\gamma(g_1,g_2)}
\end{equation}
where $\gamma:G\times G\rightarrow \bbR/\bbZ=\U(1)$. Associativity of $\U(\maH)$ requires that
\begin{align}
    \gamma(h,k)-\gamma(g h,k)+\gamma(g,hk)-\gamma(g,h)=0
\end{align}
which implies that $\gamma$ is a cocycle in the second group cohomology $H^2(G,\bbR/\bbZ)$.

If we can find a true homomorphism $\alpha:G\rightarrow \U(\maH)$ which differs from $\rho'$ only by a phase factor, then we say that $\rho$ ``lifts" to a true representation $\alpha$ of $G$.
Let the projection from $\U(\maH)$ to $\PU(\maH)$ be called $\pi$. $\rho$ lifting to $\alpha$ is equivalent to the commutation of the diagram:

\begin{equation}
    \begin{tikzcd}
        G \arrow[rd,"\rho"] \arrow[r,"\alpha"]&\U(\mathcal{H})\arrow[d,"\pi"]\\
        &\PU(\mathcal{H})
    \end{tikzcd}
\end{equation}

Such a lift is not always possible. Indeed, if the homomorphism $\alpha$ is defined such that $\rho'(g)=e^{if(g)} \alpha(g)$ for $f:G\rightarrow \bbR/\bbZ$ then we find that
\begin{equation}
    \gamma(g,h)=f(gh)-f(g)-f(h)
\end{equation}
which means that $\gamma$ is a coboundary in $H^2(G,\bbR/\bbZ)$.
Consequently, it is only possible to lift a projective representation $\rho$ of $G$ to a true representation if the map $\gamma$ lies in the trivial second group cohomology class of $G$.

It is, however, always possible to lift a projective representation to a representation of a different group $\tilde{G}$, a central extension of $G$. This is done as follows. Let $\tilde{G}:=\{(g,u)\in G\times \U(\mathcal{H})|\pi(u)=\rho(g)\}$. $\rho$ and $\pi$ are both homomorphisms so it is easy to see that $\tilde{G}$ is a group. $\tilde{G}$ is also a Lie group but we will not show that here \cite{Simms}. There is then a  short exact sequence

\begin{equation}
    \begin{tikzcd}
        0 \arrow[r]&\U(1) \arrow[r,"i"]&\tilde{G}\arrow[r,"\sigma"] &G\arrow[r]&0
    \end{tikzcd}
\end{equation}
where $i$ is shorthand for the map $\U(1)\rightarrow \U(\maH)\rightarrow \tilde{G}$ and $\sigma$ denotes projection from $\tilde{G}$ to $G$. The kernel of $\sigma$ is the set $\{(e,u)\in \tilde{G}|u\in \ker{\pi}\}$. We know $\ker{\pi}$ is precisely the image of $\U(1)$ in $\U(\maH)$ and hence the sequence is exact. This means that $\tilde{G}$ is a \textbf{one-dimensional central extension} of $G$. The diagram
\begin{equation}
\begin{tikzcd}
    \tilde{G}\arrow[r]\arrow[rd,"\rho\circ\sigma"]&\U(\maH)\arrow[d,"\pi"]\\
    & \PU(\maH)
\end{tikzcd}
\end{equation}
commutes by definition and thus $\rho$ lifts to a true unitary representation of $\tilde{G}$ always.

Hence to classify the unitary projective representations of $G$ it is sufficient to classify the unitary representations of all one-dimensional central extensions of $G$. Since all unitary representations are totally reducible, this problem reduces to classifying the \textbf{unitary irreducible representations} (UIRs) of the one-dimensional central extensions of $G$.
\subsection{The Method of Mackey}
The primary tool we will use to classify the unitary irreducible representations of the Lifshitz groups is the method of Mackey \cite{MR2069175}\cite{MR207908}\cite{MR89999}. Mackey's method allows us to find the UIRs of a Lie group $G$, which breaks up as
\begin{equation}
G=K\ltimes T
\end{equation}
where $K$ and $T$ are Lie groups. In general, the action of $K$ on $T$ is by inner automorphism. Given $\sk \in K<G$ and $\st\in T<G$, the action is
\begin{equation}
\sk\cdot \st := \sk\st\sk^{-1}
\end{equation}
We consider $\hat{T}$, the unitary dual of $T$, that is the space of UIRs of $T$, modulo equivalence as representations. The action of $K$ on $T$ induces an action of $K$ on $\hat{T}$ as follows. Let $\chi\in \hat{T}$, that is, $\chi$ is a UIR of $T$. For $\sk\in K$ and $\st\in T$ we define
\begin{equation}
[\sk\cdot \chi](\st):=\chi(\sk^{-1}\cdot \st)
\end{equation}
We call the semidirect product \textbf{regular} is there is a countable collection of $G-$invariant Borel subsets $\{E_i\}$ of $\hat{T}$ such that each $K-$orbit in $\hat{T}$ is the union of all $E_i$ which contain it. In this case, Mackey's method gives all UIRs of $G$. Given $\chi \in \hat{T}$, we define its orbit and stabiliser
\begin{equation}
\begin{split}
    \mathcal{O}_\chi &:= \{\sk\cdot \chi |\sk\in K\}\\
    K_\chi &:= \{\sk\in K| \sk\cdot \chi \cong\chi\}
\end{split}
\end{equation}
where $\cong$ denotes equivalence as representations. Let $\chi:T\rightarrow \U(V)$ be a UIR of $T$ carried by the vector space $V$. For each $\sk\in K_\chi$, the representation $\sk\cdot \chi$ is equivalent to $\chi$. That is, we may identify the carrier space of each $\sk\cdot\chi$ with $V$ and there is a homomorphism $S:K_\chi\rightarrow \U(V)$ such that 
\begin{equation}
[\sk\cdot\chi](\st)=S(\sk)^{-1}\chi(\st)S(\sk)
\end{equation}
for all $\st\in T$. Let $\rho:K_\chi\rightarrow \U(Y)$ be a UIR of $K_\chi$. Let $W = Y\otimes V$. Then we can build a UIR of $G_\chi := K_\chi\ltimes T$:
\begin{equation}
D=\rho\otimes \chi S
\end{equation}
This is a map from $G_\chi$ to $\U(W)$ which behaves as
\begin{equation}
D(\st\sk)=\rho(\sk)\chi(\st)S(\sk)
\end{equation}
for $\st\in T$ and $\sk \in K_\chi$. Since $K_\chi$ stabilises $\chi$, this is in fact a representation:
\begin{equation}
    \begin{split}
        D(\st_1 \sk_1 \st_2 \sk_2)&=D(\st_1 (\sk_1\cdot \st_2)\sk_1\sk_2)\\
        &=\rho(\sk_1 \sk_2)\chi(\st_1(\sk_1\cdot \st_2))S(\sk_1\sk_2)\\
        &=\rho(\sk_1\sk_2)\chi(\st_1)[\sk^{-1}\cdot\chi](\st_2)S(\sk_1)S(\sk_2)\\
        &=\rho(\sk_1)\rho(\sk_2)\chi(\st_1)S(\sk_1)\chi(\st_2)S(\sk_2)\\
        &=D(\st_1\sk_1) D(\st_2\sk_2)
    \end{split}
\end{equation}
We call $D$ the \textbf{inducing representation} generated by $\chi$ and $\rho$ as it induces a UIR of $G$. It turns out that elements of the same $K-$orbit in $\hat{T}$ induce equivalent representations of $G$ and hence the inducing representations are classified by the $K-$orbits in $\hat{T}$ and the representations of $K_\chi$. The UIR of $G$ induced by $D$ is defined as follows. We consider the space of \textbf{Mackey functions} $C^{\infty}_{D}(G,W)$, the space of $G_\chi-$equivariant functions from $G$ to the vector space $W$. By $G_\chi-$equivariance we mean that
\begin{equation}
F(\sg\sh)=D(\sh^{-1})F(\sg)
\end{equation}
for $\sg\in G$ and $\sh\in G_\chi$. The space $C^\infty_{D}(G,W)$ naturally carries a representation of $G$. For $\sg,\sg'\in G$, we define
\begin{equation}
[\sg\cdot F](\sg'):=F(\sg^{-1}\sg')
\end{equation}
The fact that $\sg$ acts on $\sg'$ on the right means that the $G-$action indeed sends Mackey functions to Mackey functions. We can impose a $G-$invariant inner product on $C_D^\infty(G,W)$ as follows.

Assuming $K-$orbits in $\hat{T}$ are manifolds, and by letting $T$ act trivially on $\mathcal{O}_\chi$, we  turn $\mathcal{O}_\chi=K/K_\chi$ into a homogeneous space of $G$, as $\mathcal{O}_\chi = G/G_\chi$. In particular $\pi: G\rightarrow \mathcal{O}_\chi$ is a principal $G_\chi-$bundle. We can then consider the associated  bundle 
\begin{equation}
\pi: E=G\times_D W\rightarrow \mathcal{O}_\chi
\end{equation}
which may be a Hilbert bundle if $W$ is infinite-dimensional. The spaces $C_D^\infty(G,W)$ and $\Gamma(E)$ are isomorphic as $C^\infty(\mathcal{O}_\chi)-$modules, with the isomorphism defined as follows. Let $\sigma: \mathcal{O}_\chi\rightarrow G$ be a local section. Then locally we have the map $C^\infty_D(G,W)\rightarrow \Gamma(E)$ given by
\begin{equation}
F\rightarrow [\sigma(p),F(\sigma(p))]
\end{equation}
Since the transition functions of the bundle $E$ are elements of $G_\chi$ and $F$ is $G_\chi-$equivariant, these local homomorphisms agree on the overlap of different sections to give a global map. The map is an isomorphism with inverse 
\begin{equation}
[\sigma(p),\psi(p)]\rightarrow F
\end{equation}
where $F(\sg)=D(\sg^{-1}\sigma(\pi(\sg)))\psi(\pi(\sg))$. Locally, the $G-$action on a section $[\sigma(p),\psi(p)]$ is given as
\begin{equation}
[\sg\cdot\psi](p)=D(\sh(g^{-1},p)^{-1})\psi(\sg^{-1}\cdot p)
\end{equation}
where the compensating gauge transformation $\sh \in G_\chi$ is defined by
\begin{equation}
\sg\sigma(p)=\sigma(\sg\cdot p)h(\sg,p)
\end{equation}
If there is a $G-$invariant top  form $\omega$ on $\mathcal{O}_\chi$, we can impose a $G-$invariant inner product, locally defined as
\begin{equation}\label{inny}
\langle\psi,\phi\rangle:=\int_{\mathcal{O}_\chi}\omega(p)\langle \psi(p),\phi(p)\rangle_W
\end{equation}
where $\langle\cdot,\cdot\rangle_W$ is the $G_\chi-$invariant inner product on $W$. Globally, since transition functions on $E$ are in $G_\chi$ and $\langle\cdot,\cdot\rangle_W$ is $G_\chi-$invariant, the inner product on local sections agrees on the overlaps to give an inner product on global sections. This construction thus amounts to a unitary representation of $G$, provided we restrict to those sections which are $L^2$ with respect to this inner product and take the Hilbert space completion. Sometimes a $G-$invariant top form on $\mathcal{O}_\chi$ does not exist. In this case, as long as $\mathcal{O}_\chi$ is orientable, the space of top forms is one-dimensional as a $C^\infty(\mathcal{O}_\chi)-$module, which means we can always find a quasi-invariant top form $\omega$, such that
\begin{equation}
    [\sg^* \omega](p) = f(p)\omega(p)
\end{equation}
for some function $f$, which is always positive. We then simply redefine the $G-$action on a section $[\sigma(p),\psi(p)]$ to be
\begin{equation}
    [\sg\cdot \psi](p):=\frac{1}{\sqrt{f(\sg^{-1}\cdot p)}}D(\sh(g^{-1},p)^{-1})\psi(\sg^{-1}\cdot p)
\end{equation}
and the inner product \ref{inny} is again $G-$invariant.
Two facts are stated without proof \cite{MR207908}\cite{MR89999}:
\begin{enumerate}
    \item The representation is irreducible.
    \item All UIRs of $G$ arise in this way.
\end{enumerate}
In the case where $T$ is simply-connected and abelian, $\hat{T}$ is simply the coalgebra $\tagg^*$ dual to $\tagg:=\text{Lie}(T)$. Each $\tau\in \tagg^*$ gives a one-dimensional UIR of $T$ as follows. Each $\st\in T$ is of the form $\st=\exp X$ for some $X\in \lagg$. The representation is
\begin{equation}
\chi_\tau(\exp X):=e^{i\langle \tau,X\rangle}
\end{equation}
The action of $K$ on $\hat{T}$ is simply the coadjoint action of $K$ on $\tagg^*$. 

We thus have an ``algorithm'' for generating UIRs of any group of Mackey form $G=K\ltimes T$:
\begin{enumerate}
    \item Identify the $K-$orbits in $\hat{T}$ (or $\tagg^*$ in the simply-connected abelian case). 
    \item Find the UIRs of the stabilisers of points in $\hat{T}$.
    \item Build the inducing representations $D$. If the semidirect product is regular this will induce all UIRs of $G$.
    \item Induce the UIRs as sections of associated $G_\chi-$bundles over the orbits.
\end{enumerate}
\section{The ``Building-Block'' Groups}
To understand the coadjoint orbits and UIRs of the Lifshitz groups and central extensions which break down into ``building-block'' groups, it is enough to understand the coadjoint orbits and UIRs of the building-block groups themselves. In short, if a group $G$ breaks down as a Cartesian product $G=H\times K$, then its coadjoint orbits will simply be the products of the coadjoint orbits of $H$ and $K$, and its UIRs will be the tensor products of those of $H$ and $K$. This section thus classifies the coadjoint orbits and UIRs of the building-block groups. None of these results are original. They can be found in various forms in \cite{MR2069175}\cite{MR4577534}\cite{MR207908}\cite{MR495836}\cite{MR1880691}\cite{MR1707750}\cite{MR803508}\cite{Figueroa-OFarrill:2024ocf}\cite{MR4655092}\cite{MR4611705}. Nevertheless, it is useful to put them all together with consistent notation. In some cases, the coadjoint orbits and UIRs of these building-block groups give insight into understanding those of the groups in the next section, such as $\Euc(d)$ does for $A_3^z$ and $C_3$, and the oscillator group does for $A_7$ and $C_1^1$.

\subsection{The Two-dimensional Non-Abelian Group $\Aff(1)$}

The Lie group $\text{Aff}(1)$ has as its Lie algebra the two-dimensional algebra $\mathfrak{aff}(1)$, which is spanned by generators $D$ and $H$ with commutation relation 

\begin{equation}
    [D,H]=H
\end{equation}
This algebra admits a matrix realisation 

\begin{align}
    D \rightarrow \begin{pmatrix}
        1&0\\
        0 & 0
    \end{pmatrix} && H \rightarrow \begin{pmatrix}
        0 &1\\
        0&0
    \end{pmatrix}
\end{align}
A generic group element takes the form $\sg(b,a)=e^{aH}e^{bD}$, and this is realised in matrix form as

\begin{equation}
    \sg(b,a) = \begin{pmatrix}
        e^b& a\\
        0&1
    \end{pmatrix}
\end{equation}
An algebra element is given by

\begin{equation}
    \sX(b,t) = \begin{pmatrix}
        b&t\\
        0&0
    \end{pmatrix}
\end{equation}
We denote an element of the dual to the Lie algebra $\mathfrak{aff}(1)^*$ by $\sM(\delta,E)$, with the pairing

\begin{equation}
    \langle \sM(\delta,E),\sX(b,t)\rangle = \delta b+Et
\end{equation}

\subsubsection{Coadjoint Orbits}
One easily calculates that the coadjoint action is $\Ad^*_{\sg(b,a)}\sM(\delta,E)=\sM(\delta',E')$, where

\begin{equation}
  \begin{split}
    \delta' &=\delta +e^{-b}aE\\
    E'&=e^{-b}E
  \end{split}
\end{equation}
There are therefore two classes of coadjoint orbit:

\begin{enumerate}
    \item If we choose a representative of the form $\sM(\delta_0,0)$, this is stabilised by the whole group $\Aff(1)$. These orbits are pointlike orbits defined by the equations $\delta=\delta_0$ and $E=0$. We label each of the points $\mathcal{P}_{\delta_0}$.

    \item When $E\neq 0$, we are  free to scale $E$, leaving the sign invariant. We can also move $\delta$ to $0$ by choosing a suitable $a$. We can thus choose a representative of the form $\sM(0,\pm 1)$. This is stabilised by only one element, $\sg(0,0)$. Quotienting by this single element, we get that there are two coadjoint orbits, each diffeomorphic to $\bbR^2$. These orbits are the upper and lower-half planes, defined by the inequalities $E>0$ and $E<0$ respectively. We label these orbits $\mathcal{V}^+$ and $\mathcal{V}^{-}$ respectively.

\end{enumerate}

\subsubsection{UIRs}
    To find the UIRs we split the group as $\Aff(1)= K\ltimes T$ with $K$ generated by $D$ and $T$ generated by $H$. The $K-$action on $\tagg^*$ is given as $\sk(b)\cdot \tau(E)= \tau(e^{-b}E)$. There are therefore three $K-$orbits in $\tagg^*$: a one-point orbit given by $E=0$ and two ray-like orbits for positive and negative $E$, with representatives $\tau(1)$ and $\tau(-1)$ respectively. For the pointlike orbit the stabiliser of $\tau$ is the whole of $K$, whereas the stabiliser in the ray-like cases is trivial. There are thus two different classes of UIR:

    \begin{enumerate}
        \item Discrete representations indexed by the real number $\delta_0$, where the representation space is the complex numbers. The inner product is the standard complex inner product and the group action is $\sg(b,a)\cdot \psi = e^{i\delta_0 b} \psi$. We label the representations $P_{\delta_0}$.

        \item Two continuous representations, which are the spaces of functions from $\bbR^+$ and $\bbR^-$ to $\bbC$. The first we call $V^+$. It is the space $L^2(\bbR^+,\bbC)$ with respect to the inner product
        
        \begin{equation}
               \langle\psi_1,\psi_2\rangle := \int_{\bbR^{+}}\frac{dq}{q}\bar{\psi}_1(q)\psi_2(q)
        \end{equation}
        The group action is $[\sg(b,a)\cdot\psi](q)=e^{ iaq}\psi(e^{b}q)$. The second representation, which we call $V^{-}$, is the space $L^2(\bbR^{-},\bbC)$ with respect to the inner product

         \begin{equation}
               \langle\psi_1,\psi_2\rangle := \int_{\bbR^{-}}\frac{dq}{q}\bar{\psi}_1(q)\psi_2(q)
        \end{equation}
        The group action is $[\sg(b,a)\cdot\psi](q)=e^{ iaq}\psi(e^{b}q)$.
        
    \end{enumerate}
    \subsection{The Euclidean Groups $\Euc(d)$}
The standard euclidean group of dimension $d$ is $\ISO(d)\cong\SO(d)\ltimes \bbR^d$. However, in this paper we use the term ``euclidean group'' to refer to the double cover $\Euc(d)\cong\Spin(d)\ltimes \bbR^d$, where the semidirect action is in the vector representation. The algebra $\mathfrak{iso}(d)$ is spanned by $\{L_{ab},P_c\}$ with $a,b,c\in \{1,2,...,d\}$. The commutation relations are

 \begin{equation}
 \begin{split}
     [L_{ab},L_{cd}]&=\delta_{ad}L_{bc}-\delta_{ac}L_{bd}+\delta_{bc}L_{ad}-\delta_{bd}L_{ac}\\
     [L_{ab},P_c]&=\delta_{bc}P_a-\delta_{ac}P_b
    \end{split}
 \end{equation}
A generic group element can be parametrised as $\sg(R,\bv)=e^{\bv\cdot \bp}R$, where $\bv \in \bbR^d$ and $R\in \Spin(d)$. $\Spin(d)$ acts on $\bbR^d$ via the projection to $\SO(d)$. Under this parametrisation group elements multiply as
\begin{equation}
    \sg(R_1,\bv_1)\sg(R_2,\bv_2)=\sg(R_1 R_2, \bv_1+R_1 \bv_2)
\end{equation}
We define an algebra element $\sX\in \mathfrak{iso}(d)$ by $\sX(L,\bx)=\frac{d}{ds}\sg(R(s),\bx s)|_{s=0}$. We define a dual element $\sM(\lambda,\bp)\in \mathfrak{iso}(d)^*$. $\lambda$ is an element of $\mathfrak{so}(d)^*$, which we can identify with a matrix, while $\bp$ is a vector in $\bbR^d$. The pairing is
\begin{equation}
    \langle \sM(\lambda,\bp),\sX(L,\bx)\rangle=\frac{1}{2}tr(\lambda^T L)+\bp\cdot \bx
\end{equation}
\subsubsection{Coadjoint Orbits}
When $d=2$, the coadjoint action is
\begin{equation}
    \Ad^*_{\sg(R,\bv)}\sM(\lambda,\bp)=\sM(\lambda',\bp')
\end{equation}
where
\begin{equation}
    \begin{split}
        \lambda'&=\lambda+\det(R\bp,\bv)\epsilon\\
        \bp'&=R\bp
    \end{split}
\end{equation}
Where $\epsilon$ denotes the totally antisymmetric $2\times2$ matrix:
\begin{equation}
    \epsilon=\begin{pmatrix}
        0&1\\
        -1&0
    \end{pmatrix}
\end{equation}
There are two classes of coadjoint orbit:
 \begin{enumerate}
     \item When $\bp=\boldsymbol{0}$, a generic orbit representative is $\sM(\lambda_0,\boldsymbol{0})$. The stabiliser of this point is the entire group $\Euc(2)$. The orbits are thus pointlike and given by the equations $\boldsymbol{p}=\boldsymbol{0}$ and $\lambda=\lambda_0$. We label these $\mathcal{P}_{\lambda_0}$.

     \item When $\bp\neq \boldsymbol{0}$, we rotate $\bp$ so that it lies on the x-axis: $\bp=p_0\bold{e}_1$ with $p_0>0$. A suitable choice of $\bv$ brings the representative to $\sM(0,p_0\boldsymbol{e}_1)$. Since $R$ acts via the projection to $\SO(2)$, there are two elements of $\U(1)$ which stabilise a given $\bp$: $1$ and $-1$. The stabiliser of the point $M$ is therefore $G_M = \{\sg(R,v\boldsymbol{e}_1)|R\in \mathbb{Z}_2,v\in \bbR\}\cong \mathbb{Z}_2\times \bbR$. Noting that the orbit $\mathcal{O}_M = G/G_M$, we see that $\mathcal{O}_M \cong T^*S^1$. Explicitly this is the cotangent bundle of the circle in $\bbC$ given by the equation $|\bp|=p_0$. We label these orbits $\mathcal{TS}^1_{p_0}$.
 \end{enumerate}
When $d=3$ the coadjoint action is
\begin{equation}
\Ad^*_{\sg(R,\boldsymbol{v})}\sM(\lambda,\boldsymbol{p})=\sM(\lambda',\boldsymbol{p}')
\end{equation}
where
 \begin{equation}
    \begin{split}
     \lambda'&=R\lambda R^{-1} +R\boldsymbol{p}\boldsymbol{v}^T-\boldsymbol{v}(R\boldsymbol{p})^T\\
     \boldsymbol{p}'&=R\boldsymbol{p}
     \end{split}
 \end{equation}
 The fundamental and adjoint representations of $\mathfrak{so}(3)$ are equivalent, which allows us to send $\lambda \in \mathfrak{so}(3)^*$ to $\boldsymbol{\lambda}\in \bbR^3$. Under this isomorphism the coadjoint action becomes
  \begin{equation}
  \begin{split}
      \boldsymbol{\lambda}'&=R\boldsymbol{\lambda}+\boldsymbol{v}\times  R\boldsymbol{p}\\
     \boldsymbol{p}'&=R\boldsymbol{p}
     \end{split}
 \end{equation}
 There are three classes of orbit:
 \begin{enumerate}
     \item The single point-like orbit given by $\boldsymbol{p}=\boldsymbol{\lambda}=\boldsymbol{0}$. We call this $\mathcal{P}$.

     \item If we choose a representative of the form $\sM(\boldsymbol{\lambda}\neq \boldsymbol{0},\boldsymbol{0})$, we can rotate $\boldsymbol{\lambda}$ to get a representative of the form $\sM(\lambda_0\boldsymbol{e}_1,\boldsymbol{0})$, with $\lambda_0> 0$. This point is stabilised by $G_M=\{\sg(R,\boldsymbol{v})|R\boldsymbol{e}_1=\boldsymbol{e}_1\}\cong \U(1)\ltimes \bbR^3$, where the semidirect action is as rotation in the $\boldsymbol{e}_1-$axis. Quotienting by this gives an orbit with geometry $S^2$, explicitly given by the equations $|\boldsymbol{\lambda}|=\lambda_0$ and $\boldsymbol{p}=\boldsymbol{0}$. We label such orbits $\mathcal{SP}_{\lambda_0}$

     \item When $\boldsymbol{p}\neq 0$, we can rotate $\boldsymbol{p}$ and set it equal to $p_0\boldsymbol{e}_3$ with $p_0>0$. We then choose a suitable $\boldsymbol{v}$ to bring $\boldsymbol{\lambda}$ parallel to $\boldsymbol{e}_1$ and get a coadjoint orbit representative of the form $\sM(\lambda_0 \boldsymbol{e}_1,p_0\boldsymbol{e}_1)$ with $\lambda_0 \in \bbR$. The stabiliser of this point is given by $G_M =\{\sg(R,\boldsymbol{v})|R\boldsymbol{e}_1=\boldsymbol{e}_1,\boldsymbol{v}\times \boldsymbol{e}_1=0\}\cong \U(1)\times \bbR$. Quotienting by this stabiliser gives an orbit with equations $\boldsymbol{\lambda}\cdot\bp=\lambda_0 p_0$ and $|\boldsymbol{p}|=p_0$. We can perform a change of coordinates $(\boldsymbol{\lambda},\boldsymbol{p})\rightarrow (\boldsymbol{k},\boldsymbol{p})$ where $\boldsymbol{k}= \frac{\lambda_0}{p_0}\boldsymbol{p}-\boldsymbol{\lambda}$. The equations of the orbit in these coordinates are $\boldsymbol{k}\cdot\boldsymbol{p}=0$ and $|\boldsymbol{p}|=p_0$, which exhibits this orbit as the cotangent bundle of the 2-sphere $T^*S^2$. We label these orbits $\mathcal{TS}^2_{\lambda_0,p_0}$.
 \end{enumerate}
\subsubsection{UIRs}

 $\Euc(d)$ has the natural Mackey split $\Spin(d)\ltimes \bbR^d$. The first step towards classifying the UIRs is to find the $\Spin(d)-$orbits in the dual to $\bbR^d$. The $\Spin(d)-$action on the dual is simply the restriction of the full coadjoint action given above, namely that $R\in \Spin(d)$ sends $\bp\in \bbR^d$ to $R\bp$.
 There are the following orbits:
\begin{enumerate}
    \item A point orbit at the origin given by $\bp=\boldsymbol{0}$.

    \item Orbits of the form $S^{d-1}$, given by $|\bp|=p_0$ for some $p_0>0$.
\end{enumerate}
 This gives rise to the following representations. When $d=2$ we get
 \begin{enumerate}
     \item Discrete representations carried by $\bbC$ with the standard inner product, indexed by an integer character $n$ of the group $\U(1)$. The group action is
    \begin{equation}
        \sg(R,\bv)\cdot \psi = \nu_n(R) \psi 
    \end{equation}
    where $\nu_n(e^{i\theta})=e^{in\theta}$. The associated character is $n\in \bbZ$. We label these representations $P_n$.
    
    \item Representations carried by sections of bundles over the circle $S^1$. The stabiliser of $\bp \in \bbR^2$ under the $\U(1)$ action is $\bbZ^2$ and there are two UIRs of $\bbZ_2$, the trivial representation $\rho^+: \bbZ_2\rightarrow \U(\bbC)$ and the fundamental representation $\rho^{-}:\bbZ_2\rightarrow \U(\bbC)$.  In each case, the dual element $\bp$ defines a character $\chi_{\bp}: \bbR^2\rightarrow \U(\bbC)$ defined as
    \begin{equation}
        \chi_{\bp}(\bv) = e^{i\bp\cdot\bv}
    \end{equation}
    This gives rise to two inducing representations of $\bbZ_2\times \bbR^2$: 
    \begin{align}
        D^+(R,\bv)=\rho^+(R)\chi_{\bp}(\bv) && D^{-}(R,\bv)=\rho^{-}(R)\chi_{\bp}(\bv)
    \end{align}
    The representations of $\Euc(2)$ derived from these inducing representations are respectively the spaces of sections of the associated complex line bundles
    \begin{align}
        \begin{tikzcd}
            E_{+}= \U(1)\times_{\rho^{+}}\bbC \arrow[d] \\
            S^1
        \end{tikzcd}&&
        \begin{tikzcd}
            E_{-}= \U(1)\times_{\rho^{-}}\bbC \arrow[d] \\
            S^1
        \end{tikzcd}
    \end{align}
    These bundles are respectively the trivial complex line bundle over $S^1$ and the complex M\"obius bundle. We thus get two different UIRs, the trivial and M\"obius representations. In each case the inner product is
    \begin{equation}\label{iner}
        \langle \psi_1,\psi_2\rangle = \int_{S^1}\bar{\psi}_1(\theta)\psi_2(\theta)d\theta
    \end{equation}
    Let $\bp=(p_0\cos \theta, p_0\sin\theta)^T$. The group action is as follows. If $R\in \U(1)$, then
    \begin{equation}
        [\sg(R,\bv)\cdot \psi](\bp)=
            Q(R,\bp)e^{i\bp\cdot\bv}\psi(R^{-1}\bp)
    \end{equation}
    where $Q(R,\bp)$ accounts for the possible twist in the bundle. It is defined as follows. Let $R=e^{i\alpha}$. Then
     \begin{equation}\label{Twist}
        Q(R,\bp)= \begin{cases}
            1\text{ if there is no twist (the bundle is trivial), or: }\\
            1\text{ if } \sin(\theta-2\alpha) >0 \text{ or } \cos(\theta-2\alpha) = 1\\
            -1\text{ if } \sin(\theta-2\alpha)< 0 \text{ or } \cos(\theta-2\alpha) = -1
        \end{cases}
    \end{equation}
    Each of these representations has the associated character $p_0\in \bbR^+$. We label the trivial and twisted representations $TS_{p_0}^+$ and $TS^{-}_{p_0}$ respectively.
 \end{enumerate}
    When $d=3$ we get the representations
    \begin{enumerate}
        \item Discrete representations, one for each UIR of $\SU(2)$, carried by $\bbC^{n+1}$ with the standard inner product. Let $\rho_n:\SU(2)\rightarrow \bbC^{n+1}$ be the standard reps of $\SU(2)$. Then the group action is
        \begin{equation}
            \sg(R,\bv)\cdot\psi=\rho_n(R)\psi
        \end{equation}
        The associated character is $n\in \bbN_0$. We label these representations $P_{n}$.

        \item Representations given by sections of bundles over the two-sphere $S^2$. The stabiliser of the point $\bp \in \bbR^3$ under the action of $\SU(2)$ is a $\U(1)$ subgroup, whose UIRs are carried by $\bbC$ and characterised by the integers according to $\nu_n(e^{i\theta})=e^{in\theta}$. We can choose a representative $\bp=p_0\boldsymbol{e}_3$, with $p_0>0$, on the two sphere to arrive at an inducing representation of $\U(1)\ltimes \bbR^3$:
        \begin{equation}
            D_{n,p_0}(R,\boldsymbol{v})=\nu_n(R)e^{i\bp\cdot \boldsymbol{v}}
        \end{equation}
        Each inducing representation gives rise to an associated bundle 
        \begin{equation}
            \begin{tikzcd}
                \Euc(3)\times_{D_{n,p_0}}\bbC \arrow[d]\\
                S^2
            \end{tikzcd}
        \end{equation}
         This bundle is the tensor product of two bundles over the two-sphere, the trivial bundle $S^2\times \bbC$ and the homogeneous line bundle of degree $n$:
        \begin{equation}
            \mathcal{O}(-n)\rightarrow S^2
        \end{equation}
        The induced representations of $\Euc(3)$ are thus carried by sections of $\mathcal{O}(-n)$, which are $L^2$ with respect to the inner product
        \begin{equation}
            \langle \psi_1,\psi_2\rangle= \int_{S^2}\bar{\psi}_1\psi_2 d\Omega
        \end{equation}
        where $d\Omega$ is the standard volume form on $S^2$, given in stereographic coordinates $\zeta$ as
        \begin{equation}\label{vol}
            d\Omega=\frac{2i d\zeta \wedge d\bar{\zeta}}{(1+|\zeta|^2)^2}
        \end{equation}
        In local stereographic coordinates $\zeta$ the group action is as follows. Let
        \begin{equation}
            R=\begin{pmatrix}
                a & b\\
                -\bar{b} & \bar{a}
            \end{pmatrix}
        \end{equation}
        with $|a|^2+|b|^2=1$ denote an element of $\SU(2)$. Then 
        \begin{equation}
                [\sg(R,\boldsymbol{v})\cdot\psi](\zeta)= (\frac{\bar{a}-b\zeta}{|\bar{a}-b\zeta|})^{-n}e^{i\boldsymbol{p}(\zeta)\cdot\boldsymbol{v}}\psi(\frac{a\zeta+\bar{b}}{-b\zeta+\bar{a}})
     \end{equation}
    where $\bp(\zeta)$ embeds the point $\zeta$ on the two sphere with radius $p_0$ into $\bbR^3$. For more on the derivation of this group action see appendix \ref{app1}. The associated characters are $n\in \bbZ$ and $p_0 \in \bbR^+$. We label these representations $TS^2_{n,p_0}$.
    \end{enumerate}
\subsection{The groups $\Spin(d+1)$}

$\Spin(d+1)$ is the simply-connected group with algebra $\mathfrak{so}(d+1)$, provided $d>1$. It is well known that $\Spin(3)\cong \SU(2)$ and $\Spin(4)\cong \SU(2)\times \SU(2)$. It is therefore enough to find the coadjoint orbits and UIRs of $SU(2)$. These are well-known. $\SU(2)$ is semisimple so its adjoint and coadjoint orbits are in correspondence. The adjoint orbits are 2-spheres, each given by a radius $\lambda_0$. In the degenerate case where $\lambda_0=0$ there is a pointlike orbit. The UIRs arise as quantisations of these orbits, or alternatively via the standard construction with roots and weights. There is one for each whole number $n$. The corresponding representation space is $\bbC^{n+1}$.

\subsection{The groups $\Spin(d,1)$}
These groups are the simply-connected groups with algebra $\mathfrak{so}(d,1)$. It is well-known that $\Spin(2,1)\cong \SL(2,\bbR)$ and $\Spin(3,1)\cong \SL(2,\bbC)$. Let us first deal with the group $\SL(2,\bbC)$ as the fact that $\bbC$ is algebraically closed makes our task easier. 

\subsubsection{Coadjoint orbits of $SL(2,\bbC)$}
The group $\SL(2,\bbC)$ is semisimple so we need only calculate adjoint orbits. The algebra $\mathfrak{sl}(2,\bbC)=\Lie(\SL(2,\bbC))$ is the vector space consisting of all traceless anti-hermitian matrices with the commutator as the Lie bracket. The adjoint action is simply conjugation of these matrices by elements of $\SL(2,\bbC)$. By conjugating with elements of $\GL(2,\bbC)$ we can get matrices in $\mathfrak{sl}(2,\bbC)$ into Jordan normal form. Since conjugating by $\frac{1}{\sqrt{\det K}}K \in \SL(2,\bbC)$ has the same effect as conjugating by $K \in \GL(2,\bbC)$, each Jordan normal form corresponds to an adjoint orbit of $\SL(2,\bbC)$. $\mathfrak{sl}(2,\bbC)$ admits the following Jordan normal forms:

\begin{equation}
    \begin{pmatrix}
        0&0\\
        0&0
    \end{pmatrix} \qquad 
    \begin{pmatrix}
        m & 0\\
        0 & -m
    \end{pmatrix} \qquad
    \begin{pmatrix}
        0 & 1 \\
        0 & 0
    \end{pmatrix}
\end{equation} 
where $m\in \bbC$. Note that the matrices 
\begin{equation}
    \begin{pmatrix}
        m & 0\\
        0 & -m
    \end{pmatrix} \qquad
    \begin{pmatrix}
        -m & 0\\
        0 & m
    \end{pmatrix} 
\end{equation}
are equivalent up to the permutation of Jordan blocks and therefore give rise to the same coadjoint orbit. Let $\lambda$ denote a matrix in $\mathfrak{sl}(2,\bbC)$. We get the following adjoint orbits

\begin{enumerate}
    \item The matrix $\begin{pmatrix}
        0&0\\
        0&0
    \end{pmatrix}$ gives rise to its own one-point orbit with equation $\lambda =0$. We label it $\mathcal{P}$.
    \item The stabiliser of $\begin{pmatrix}
        m &0\\
        0&-m
    \end{pmatrix}$ under the adjoint action is the maximal torus in $\SL(2,\bbC)$, namely the group $T=\{\begin{pmatrix}
        w &0\\
        0& w^{-1}
    \end{pmatrix}|w \in \bbC\backslash\{0\}\}$. The orbit is thus $\SL(2,\bbC)/T$, which has the geometry $S^2\times \bbR^2$. The orbit equation is $\det(\lambda)=-m^2$ with $m\neq 0$. We label these orbits $\mathcal{U}_{m}$.

    \item The stabiliser of $\begin{pmatrix}
        0&1\\
        0&0
    \end{pmatrix}$ under the adjoint action is the  subgroup $\bbZ_2\times \bbC=\{\begin{pmatrix}
        \pm1 &z\\
        0& \pm1
    \end{pmatrix}|z\in \bbC\}$. The orbit is $\SL(2,\bbC)/(\bbZ_2\times \bbC)$ with geometry $ \mathbb{RP}^3\times \bbR^+$. The orbit relations are $\det(\lambda)=0$ and $\lambda\neq 0$. We label these orbits $\mathcal{U}_{0}$.
\end{enumerate}

\subsubsection{UIRs of $\SL(2,\bbC)$}

There are three classes of UIR \cite{MR1880691}:
\begin{enumerate}
    \item The trivial representation 

    \item The principal series. The representation space is locally the space of functions $F: \bbC \rightarrow \bbC $, which are $L^2$ with respect to the inner product:

    \begin{equation}
        \langle F_1, F_2 \rangle := \int_\bbC \bar{F_1}(\zeta)F_2(\zeta)d\zeta \wedge d\bar{\zeta}
    \end{equation}
    
    The group action is 

    \begin{equation}
        [\begin{pmatrix}
            a & b\\
            c & d
        \end{pmatrix}\cdot F](\zeta)=|-b\zeta+d|^{-2-iq}(\frac{-b\zeta +d}{|-b\zeta+d|})^{-n}F(\frac{a\zeta-c}{-b\zeta+d})
    \end{equation}
    The associated characters are $n \in \bbN_0$ and $q\in \bbR$. We label these representations $U_{n,q}$. Note that $U_{0,q}$ is unitarily equivalent to $U_{0,-q}$.
    
    \item The complementary series. The representation space is locally the space of functions $F: \bbC \rightarrow \bbC$ which are $L^2$ with respect to the inner product

    \begin{equation}
    \langle F_1, F_2 \rangle := \int_\bbC \int_\bbC |\zeta_1-\zeta_2|^{-2+w}\bar{F}(\zeta_1)F(\zeta_2)d\zeta_1 \wedge d\bar{\zeta}_1 \wedge d\zeta_2 \wedge d\bar{\zeta}_2
    \end{equation}
    The group action is
    \begin{equation}
       [\begin{pmatrix}
           a & b\\
           c & d
       \end{pmatrix}\cdot F](\zeta)=|-b\zeta+d|^{-2-w}F(\frac{a\zeta-c}{-b\zeta+d})
    \end{equation}
    The associated character is $w\in (0,2)$. We label these representations $C_w$.
\end{enumerate}

\subsubsection{Coadjoint orbits of $\SL(2,\bbR)$}

The group $\SL(2,\bbR)$ is once again semisimple so we need only consider the adjoint orbits. The conjugate action of $\GL(2,\bbR)$ on elements of $\mathfrak{sl}(2,\bbR)$ gives rise to the real Jordan normal forms of $2\times 2$ matrices. Since $\bbR$ is not algebraically closed, the conjugate action of $\GL(2,\bbR)$ on $\mathfrak{sl}(2,\bbR)$ elements cannot be reduced to that of $\SL(2,\bbR)$. The most we can do is reduce the action of $K \in \GL(2,\bbR)$ to $\frac{1}{\sqrt{|\det K|}}K$. This matrix lives in $\GL(2,\bbR)^{\pm}=\{W \in \GL(2,\bbR)|\det(W)=\pm 1\}$, which is essentially two copies of $\SL(2,\bbR)$. This means that the adjoint action of $\SL(2,\bbR)$ puts matrices into real Jordan normal form modulo the conjugate action of $\begin{pmatrix}
    1 &0\\
    0 &-1
\end{pmatrix}$, which flips between the positive and negative determinant subgroups of $\GL(2,\bbR)^\pm$. The matrix algebra $\mathfrak{sl}(2,\bbR)$ admits the following real Jordan normal forms:

\begin{equation}
    \begin{pmatrix}
        0&0\\
        0&0
    \end{pmatrix} \qquad 
    \begin{pmatrix}
        m & 0\\
        0 & -m
    \end{pmatrix} \qquad
    \begin{pmatrix}
        0 & -m\\
        m & 0
    \end{pmatrix} \qquad
    \begin{pmatrix}
        0 & 1 \\
        0 & 0
    \end{pmatrix}
\end{equation} 
where $m$ is a positive real number. When we account for the action of $\begin{pmatrix}
    1 &0\\
    0& -1
\end{pmatrix}$, we see that $\SL(2,\bbR)$ admits the following adjoint orbit representatives:

\begin{equation}
    \begin{split}
    &\begin{pmatrix}
        0&0\\
        0&0
    \end{pmatrix} \qquad 
    \begin{pmatrix}
        m & 0\\
        0 & -m
    \end{pmatrix} \qquad
    \begin{pmatrix}
        0 & -m\\
        m & 0
    \end{pmatrix} \\
    &\begin{pmatrix}
        0 & m\\
        -m & 0
    \end{pmatrix} \qquad
    \begin{pmatrix}
        0 & 1 \\
        0 & 0
    \end{pmatrix} \qquad
    \begin{pmatrix}
        0 & -1 \\
        0 & 0
    \end{pmatrix}
    \end{split}
\end{equation}
    We get the following adjoint orbits:
    \begin{enumerate}
        \item The matrix $\begin{pmatrix}
            0 & 0\\
            0& 0
        \end{pmatrix}$ gives rise to its own one-point orbit, which we call $\mathcal{P}$.

        \item The stabiliser of $\begin{pmatrix}
            m & 0\\
            0 & -m
        \end{pmatrix}$ is the maximal torus $T=\{\begin{pmatrix}
            a & 0\\
            0 & a^{-1}
        \end{pmatrix}|a\in \bbR\backslash\{0\}\}$. The orbit is thus $\SL(2,\bbR)/T$, which has the geometry $S^1\times \bbR$. The orbit equation is $\det(\lambda)=-m^2$ and $m>0$. We call these orbits $\mathcal{U}_m$.

        \item The two representatives $\begin{pmatrix}
            0 & -m\\
            m & 0
        \end{pmatrix}$ and $\begin{pmatrix}
            0 & m\\
            -m & 0
        \end{pmatrix}$ have the same stabiliser, namely the maximal compact subgroup $\SO(2)$ of $\SL(2,\bbR)$. This gives rise to orbits with geometry $\SL(2,\bbR)/\SO(2) \cong \bbR^2$. The equations are $\det(\lambda)=m^2$, $m>0$ and $\lambda_{12}<0$ or $\lambda_{12}>0$ respectively. We label these orbits $\mathcal{H}^\pm_m$.

        \item The representatives $\begin{pmatrix}
            0 &1\\
            0&0
        \end{pmatrix}$ and $\begin{pmatrix}
            0& -1\\
            0&0
        \end{pmatrix}$ have the same stabiliser, $\bbZ_2 \times \bbR = \{\begin{pmatrix}
            \pm 1& a\\
            0 & \pm 1
        \end{pmatrix}|a \in \bbR\}$. This gives rise to orbits with geometry $\SL(2,\bbR)/(\bbZ_2\times \bbR) \cong S^1 \times \bbR^\pm$. The equations are $\det(\lambda)=0$, $\lambda_{12}<0$ or $\lambda_{12}>0$ respectively. We label these $\mathcal{H}_0^\pm$
    \end{enumerate}

\subsubsection{UIRs of $\SL(2,\bbR)$}

There are 5 classes of UIR \cite{MR1880691}:
\begin{enumerate}
    \item The trivial representation, which we label P.

     \item The positive and negative principal series, called $U^+_{q}$ and $U^{-}_q$ respectively, consisting of all $F:\bbR\rightarrow \bbC$ which are $L^2$ with respect to
    \begin{equation}
        \langle F_1,F_2\rangle=\int_\bbR \bar{F}_1(x)F_2(x)dx
    \end{equation}
    The group action is
    \begin{equation}
        [\begin{pmatrix}
            a & b\\
            c & d
        \end{pmatrix}\cdot F](x)=\begin{cases}
            |-bx+d|^{-1-iq}F(\frac{ax-c}{-bx+d}) \text{ if }U^+_q\\
            \sgn(-bx+d)|-bx+d|^{-1-iq}F(\frac{ax-c}{-bx+d}) \text{ if }U^{-}_q\\
        \end{cases}
    \end{equation}
    The associated character is $q\in \bbR$. All these representations are irreducible, with the exception of $U^{-}_0$, which breaks down as
    \begin{equation}
        U^-_0=H^+_0 +H^-_0
    \end{equation}

    \item The holomorphic and antiholomorphic discrete series. The holomorphic discrete series consists of all $F: \bbC \rightarrow \bbC$ holomorphic on $Im(\zeta) > 0$, which are $L^2$ with respect to

    \begin{equation}
        \langle F_1, F_2 \rangle = \int_{Im(\zeta)>0}\bar{F}_1(\zeta)F_2(\zeta)y^{n-2}dx\wedge dy
    \end{equation}

    where $\zeta =x+iy$ and $n>2$ is an integer. The action is 
    \begin{align}
        [\begin{pmatrix}
            a & b\\
            c & d
        \end{pmatrix}\cdot F](\zeta)=(-b\zeta+d)^{-n}F(\frac{a\zeta-c}{-b\zeta+d)})
    \end{align}
    The antiholomorphic series consists of antiholomorphic functions on the upper half plane with the same inner product. The action is
    \begin{equation}
        [\begin{pmatrix}
            a & b\\
            c & d
        \end{pmatrix} \cdot F](\zeta)=(-b\bar{\zeta}+d)^{-n}F(\frac{a\zeta-c}{-b\zeta+d})
    \end{equation}
    The holomorphic and antiholomorphic discrete series have the associated integer character $n>2$. We label them $H^\pm_n$.
    \item The holomorphic and antiholomorphic limit discrete series. These are holomorphic and antiholomorphic functions $L^2$ with respect to the inner product induced from the norm
    \begin{equation}
        ||F||^2=\sup_{y>0}\int_{-\infty}^{\infty}|F(x+iy)|^2 dx
    \end{equation}
    The action for the holomorphic limit discrete series is
    \begin{equation}
        [\begin{pmatrix}
            a & b\\
            c & d
        \end{pmatrix}\cdot F](\zeta)=(-b\zeta+d)^{-1}F(\frac{a\zeta-c}{-b\zeta+d})
    \end{equation}
    while the action for the antiholomorphic limit discrete series is
     \begin{equation}
        [\begin{pmatrix}
            a & b\\
            c & d
        \end{pmatrix}\cdot F](\zeta)=(-b\bar{\zeta}+d)^{-1}F(\frac{a\zeta-c}{-b\zeta+d})
    \end{equation}
    We call these representations $H^\pm_0$.

    \item The complementary series consisting of all $F:\bbR\rightarrow \bbC$ which are $L^2$ with respect to
    \begin{equation}
        \langle F_1,F_2\rangle=\int_\bbR \int_\bbR |x-y|^{u-1}\bar{F}_1(x)F_2(y)dx\wedge dy
    \end{equation}
    The group action is
    \begin{equation}
        [\begin{pmatrix}
            a & b\\
            c & d
        \end{pmatrix}\cdot F](x)=|-bx+d|^{-1-u}F(\frac{ax-c}{-bx+d})
    \end{equation}
    The associated character is $u\in (0,1)$. We call these representations $C_u$.
\end{enumerate}
 \subsection{The Heisenberg Group $\N$}

    This is a three-dimensional group whose algebra is spanned by $\{P_1,P_2,Z\}$ with commutation relation

    \begin{equation}
        [P_1,P_2]=Z
    \end{equation}
    This algebra has a matrix representation given by

    \begin{equation}
     P_1\rightarrow \begin{pmatrix}
            0&1&0\\
            0&0&0\\
            0&0&0
        \end{pmatrix} \qquad
        P_2\rightarrow \begin{pmatrix}
            0&0&0\\
            0&0&1\\
            0&0&0
        \end{pmatrix} \qquad
        Z\rightarrow \begin{pmatrix}
            0&0&1\\
            0&0&0\\
            0&0&0
        \end{pmatrix}
    \end{equation}
    We choose a group parametrisation $\sg(v_1,v_2,\gamma)=e^{v_1 P_1+v_2 P_2}e^{(\gamma-\frac{1}{2}v_1 v_2)Z}$, so that $\sg$ looks like

    \begin{equation}
        \sg(v_1,v_2,\gamma)\rightarrow\begin{pmatrix}
            1&v_1&\gamma\\
            0&1&v_2\\
            0&0&1
        \end{pmatrix}
    \end{equation}
    in the matrix representation. We can alternatively write the parametrisation with the vector $\bv=(p_1,p_2)^T$ as $\sg(\bv,\gamma)=\sg(R,\boldsymbol{v},\gamma)=e^{\boldsymbol{v}\cdot\boldsymbol{P}} e^{(\gamma-\frac{1}{4}\boldsymbol{v}^T B \boldsymbol{v})Z}$, where 
    \begin{equation}
        B=\begin{pmatrix}
            0&1\\
            1&0
        \end{pmatrix}
    \end{equation}
    This parametrisation using a vector as opposed to two scalars will be useful when we get to the oscillator group, so we adopt it here. 
    A generic Lie algebra element is given by $\sX(\bx,u)=\frac{d}{ds}\sg(\bx s,us)|_{s=0}$. A dual element is given by $\sM(\bp,m)$, with pairing
    \begin{equation}
        \langle \sM(\bp,m),\sX(\bx,u)\rangle = \bp\cdot \bv+m u
    \end{equation}
   
\subsubsection{Coadjoint Orbits}
The coadjoint action is
\begin{equation}
    \Ad^*_{\sg(\bv,\gamma)}\sM(\bp,m)=\sM(\bp+\epsilon\bv,m)
\end{equation}
There are two classes of coadjoint orbit.
\begin{enumerate}
    \item Choosing a representative of the form $\sM(\bp_0,0)$, we see it is stabilised by the entire Heisenberg group $N$. These points therefore define their own one-point orbits given by the equations $\bp=\bp_0$ and $m=0$. We label these points $\mathcal{P}_{\bp_0}$.

    \item If we choose a representative with non-zero $m$, $\sM(\boldsymbol{0},m_0\neq0)$, we find that its stabiliser is $G_{\sM}=\{\sg(\boldsymbol{0},\gamma)\}\cong \bbR$. The orbits are hyperplanes given by the single equation $m=m_0$. We label the hyperplanes $\mathcal{V}_{m_0}$.
\end{enumerate}

\subsubsection{UIRs}
The Heisenberg group $\N$ has the Mackey split $K\ltimes T$ where $K$ is generated by $P_1$, and $T$ is generated by $P_2$ and $Z$. The $K-$action on the dual $\tagg^*$ is $k(v_1)\cdot \tau(p_2,m) =\tau(p_2-mv_1,m)$. There are therefore two types of $K-$orbit: those with representative $\tau(p,0)$, which are pointlike, and those with representative $\tau(0,m_0\neq 0)$, which are diffeomorphic to $\bbR$. In the first case the stabiliser is the whole of $K$ and in the second it is trivial. We thus get two classes of UIR:

 \begin{enumerate}
     \item Discrete representations carried by $\bbC$ with the standard inner product. The central element acts trivially and the group action is
     \begin{equation}
         \sg(\bv,\gamma)\cdot \psi = e^{i\bp_0\cdot \bv}\psi
     \end{equation}
     The associated character is $\bp_0\in \bbR^2$. We call these representations $P_{\bp_0}$.

     \item Continuous representations carried by the space of functions $\psi:\bbR\rightarrow \bbC$, which are $L^2$ with respect to the inner product
     \begin{equation*}
         \langle \psi_1,\psi_2\rangle =\int_\bbR dq \bar{\psi}_1(q)\psi_2(q)
     \end{equation*}

     The group action is 
     \begin{equation}
         [\sg(\bv,\gamma)\cdot\psi](q)=e^{i(m_0\gamma+q v_2)}\psi(q+m_0 v_1)
     \end{equation}
     The associated character is $m_0\in \bbR\backslash\{0\}$. We label these representations $V_{m_0}$.
 \end{enumerate}

 \subsection{The Oscillator or Diamond Group $\mathcal{D}$}

The oscillator group is $\U(1)\ltimes \N$. The $\U(1)$ group has algebra generator $L$, which acts on the Heisenberg algebra as
\begin{align}
    [L,P_1]&=P_2\\
    [L,P_2]&=-P_1
\end{align}
In line with our parametrisation of the Heisenberg group, we denote a group element of the oscillator group by 
\begin{equation}
\sg(R,\boldsymbol{v},\gamma)=e^{\boldsymbol{v}\cdot\boldsymbol{P}}R e^{(\gamma-\frac{1}{4}\boldsymbol{v}^T B \boldsymbol{v})Z}
\end{equation}
where 
\begin{equation}
B=\begin{pmatrix}
    0&1\\
    1&0
\end{pmatrix}
\end{equation}
A generic algebra element is given by $\sX(L,\boldsymbol{x},u)=\frac{d}{ds}\sg(R(s),\boldsymbol{x}s,us)|_{s=0}$. A dual element is given by $\sM(\lambda,\boldsymbol{p},m)$ with pairing
\begin{equation}
    \langle \sM(\lambda,\boldsymbol{p},m),\sX(L,\boldsymbol{x},q)\rangle=\frac{1}{2}tr(\lambda^T L)+\boldsymbol{p}\cdot \boldsymbol{x}+mq
\end{equation}
\subsubsection{Coadjoint Orbits}
The coadjoint action is
\begin{equation}
\Ad^*_{\sg(R,\boldsymbol{v},\gamma)}\sM(\lambda,\boldsymbol{p},m)=\sM(\lambda',\boldsymbol{p}',m')
\end{equation}
where
\begin{equation}
\begin{split}
    \lambda'&=\lambda+\det(R\boldsymbol{p},\boldsymbol{v})\epsilon+\frac{1}{2}m|\boldsymbol{v}|^2\epsilon\\
    \boldsymbol{p}'&=R\boldsymbol{p}+m\epsilon\boldsymbol{v}\\
    m'&=m
    \end{split}
\end{equation}
Recall that $\epsilon$ is the totally antisymmetric $2\times2$ matrix:
\begin{equation}
    \epsilon=\begin{pmatrix}
        0&1\\
        -1&0
    \end{pmatrix}
\end{equation}
There are three classes of coadjoint orbit.
\begin{enumerate}
    \item Pointlike orbits given by the equations $\lambda=\lambda_0$, $\bp=\boldsymbol{0}$ and $m=0$. We label these $\mathcal{P}_{\lambda_0}$.

    \item Orbits where $m=0$ and $\bp\neq \boldsymbol{0}$. We can choose a representative $\sM(0,p_0\boldsymbol{e}_1,0)$ where $p_0>0$. The orbit of this point has the geometry $T^*S^1$, given by the equations $|\bp|=p_0$ and $m=0$. We label these $\mathcal{TS}^1_{p_0}$.

    \item When $m\neq 0$, we can choose a representative $\sM(\lambda_0,\boldsymbol{0},m_0)$. This has an orbit diffeomorphic to $\bbR^2$, given by the equations $m=m_0\neq 0$ and
    \begin{equation}
        \lambda=\lambda_0+\frac{|\bp|^2}{2m}\epsilon
    \end{equation}
    We label these orbits $\mathcal{G}_{\lambda_0,m_0}$.
\end{enumerate}
\subsubsection{UIRs}
We apply Mackey with the split $\mathcal{D}=\U(1)\ltimes \N$. The normal subgroup is not abelian for the first time in this paper. We therefore have to consider $\U(1)-$orbits in $\hat{\N}$, the unitary dual of $\N$. $\hat{\N}$ is the space of all UIRs of $\N$ up to equivalence. There are two classes of UIR of $\N$. The discrete representations are carried by $\bbC$ and specified by the character $\bp_0\in \bbR^2$. The continuous representations are specified by the non-zero character $m_0$. Letting $(\bp,m_0)$ be coordinates on $\bbR^3$, $\hat{\N}$ is given by the plane union the punctured vertical line: $\hat{\N}=\{(\bp,0)\}\sqcup\{(\boldsymbol{0},m_0)|m_0\neq 0\}$: 

    \begin{figure}[H]
    \centering
    \caption{The unitary dual of the Heisenberg Group}
    \begin{tikzpicture}[x=0.7cm,y=0.7cm]
    \coordinate (O) at (0,0);
    \coordinate (A) at (6,2); 
    \coordinate (B) at (4,-2); 
    \coordinate (C) at (-6,-2); 
    \coordinate (D) at (-4,2);
    \coordinate (P1) at (0,4); 
    \coordinate (P2) at (0,-2); 
    \coordinate (P3) at (0,-4);
    \fill [blue!10!white] (A) -- (B) -- (C) -- (D) -- cycle;
    \draw [thick,black] (O)--(P1);
    \draw [thick, loosely dotted, black] (O)--(P2);
    \draw [thick,black] (P2)--(P3);
    \filldraw [color=red!50!black,fill=red!10!white] (O) circle (2pt);
    \node at (0.5,3.7) {$m_0$};
    \node at (5.6,0) {$p_1$};
    \node at (1.7,2.5) {$p_2$};
    \end{tikzpicture}
    \caption{The discrete representations are indexed by points $(p_1,p_2,0)$ on the flat plane while the continuous representations are indexed by points $(0,0,m_0)$ on the vertical.}
    \end{figure}
The action of $\U(1)$ on on representation $\chi$ in $\hat{N}$ is given as 
\begin{equation}
[R\cdot \chi](\sg(\bv,\gamma))=\chi(R^{-1}\cdot\sg(\bv,\gamma))
\end{equation}
for $R\in \U(1)$ and $\sg(\bv,\gamma)\in \N$. It is easily computed that
\begin{equation}
R^{-1}\cdot \sg(\bv,\gamma)=\sg(R^{-1}\boldsymbol{v},\gamma-\frac{1}{4}\boldsymbol{v}^T B \boldsymbol{v}+\frac{1}{4}\boldsymbol{v}^T RB R^{-1} \boldsymbol{v})
\end{equation}
For representations in the plane, the character $\bp_0$ changes as $\bp_0\rightarrow R\bp_0$. For representations on the punctured vertical, the character $m_0$ doesn't change.
The $\U(1)$ orbits in $\hat{\N}$ are thus
\begin{enumerate}
    \item A point orbit at the origin $(\boldsymbol{0},0)$.

    \item $S^1$ orbits in the plane.

    \item Points on the punctured vertical.
\end{enumerate}
These give rise to the following UIRs of $D$.
\begin{enumerate}
    \item Discrete representations carried by $\bbC$ with the standard inner product. The group action is
    \begin{equation}
        \sg(R,\bv,\gamma)\cdot \psi = \nu_n(R)\psi
    \end{equation}
    The associated character is $n\in \bbZ$. We call these representations $P_n$.

    \item Continuous representations, given by sections of bundles over circles in the plane. These are the familiar trivial and M\"obius bundles, with the same inner product on $S^1$ given by equation \ref{iner}. The group action is
    \begin{equation}
        [\sg(R,\bv,\gamma)\cdot \psi](\bp)=Q(R,\bp)e^{i\bp\cdot \bv}\psi(R^{-1}\bp)
    \end{equation}
    where $Q(R,\bp)$ is as in equation \ref{Twist}. The associated character is $p_0\in \bbR^+$. We label these representations $TS^\pm_{p_0}$. \footnote{Note that it is no accident that the first two coadjoint orbits and UIRs of the oscillator group exactly match those of $\Euc(2)$. These coadjoint orbits and UIRs have $m=0$ and essentially correspond to forgetting the central element $Z$ in the oscillator algebra, reducing it to the algebra $\mathfrak{iso}(2)$.}
    
    \item Representations induced by representations of $\N$ lying on the punctured verticle line in $\hat{\N}$, with $m_0\neq 0$ (that is, where the central element doesn't act trivially). We have to investigate how the continuous Heisenberg representations change under the $\U(1)$ action. The continuous Heisenberg representations go as
    \begin{equation}
    [\chi(\sg(v_1,v_2,\gamma))\cdot \psi](q)=e^{i(m_0\gamma+qv_2)}\psi(q+m_0v_1)
    \end{equation}
    Under the $\U(1)$ action, $\chi$ becomes $\chi'=\chi\circ R^{-1}$. $R$ is a $\U(1)$ element so we can parametrise its action on $\bbR^2$ as
    \begin{equation}
    R\boldsymbol{v}=\begin{pmatrix}
    M & -N\\
    N & M
    \end{pmatrix}\boldsymbol{v}
    \end{equation}
    where $M=\cos(2\theta)$ and $N=\sin{2\theta}$. We thus get the following group action for $\chi'$: 
    \begin{multline}
    [\chi'(\sg(v_1,v_2,\gamma))\cdot\psi](q) = e^{im_0(\gamma-N^2v_1v_2+MN\frac{v_2^2-v_1^2}{2})}e^{iq(-Nv_1+Mv_2)} \psi(q+m_0(Mv_1+Nv_2)).
    \end{multline}
    Note that the central element $\gamma$ still acts according to the character $m_0$. This means the representation $\chi'$ must be equivalent to the representation $\chi$. It thus remains to find a unitary operator $S(R)$ such that $\chi'=S^{-1}\chi S$. It turns out $S$ can be expressed as an integral transform:
    \begin{equation}\label{Mehler}
    S(R)\psi(q) = A(R)\int_{-\infty}^{\infty}e^{\frac{i}{2m_0} \cot(2\theta)(q^2+k^2)-\frac{iqk}{m_0}\csc(2\theta)}\psi(k)dk
    \end{equation}
    where $A(R)$ is a factor chosen so that $S(1)$ is the identity. See appendix \ref{app2} for the derivation of this integral transform. 
    The induced representation we get is thus carried by the square-integrable functions $\bbR\rightarrow \bbC$ with respect to the inner product:
    \begin{equation}
        \langle \psi_1,\psi_2\rangle = \int_\bbR dq \bar{\psi}_1(q)\psi_2(q)
    \end{equation}
    The group action is
    \begin{equation}
        [\sg(R,\boldsymbol{v},\gamma)\cdot\psi](q)= \nu_n(R) e^{i(m_0\gamma+qv_2)}[S(R)\psi](q+m_0 v_1)
    \end{equation}
    We label these representations $G_{n,m_0}$. 

\end{enumerate}

\subsection{Summary of Results}\label{results1}

Table \ref{tab3} lists the bulding-block groups and their coadjoint orbits. Each orbit is specified by conditions, which take the form of equations or inequalities between the various coordinates on the coalgebra $\lagg^*$ of each group $G$. The orbit geometry is also stated up to diffeomorphism, as is the dimension of each orbit. Every orbit is even dimensional, which is to be expected for symplectic manifolds. The naming of the orbits follows a convention where the calligraphic capital letters refer to the geometry of the orbit while the subscripts denote the value of the Casimirs, which are constant on the orbit. The calligraphic capital letter encodes not just the intrinsic geometry of the orbit but also how it is embedded in the coalgebra. For example, the orbit $\mathcal{V}_{m_0}$ of the Heisenberg group and the orbit $\mathcal{G}_{\lambda_0,m_0}$ of the oscillator group are both diffeomorphic to $\bbR^2$ but have very different shapes inside their respective coalgebras. The orbit $\mathcal{V}_{m_0}$ is a flat plane while $\mathcal{G}_{\lambda_0,m_0}$ is a paraboloid defined by the condition
\begin{equation}\label{osc}
    \lambda=\lambda_0+\frac{|\bp|^2}{2m_0}\epsilon
\end{equation}
The calligraphic letter $\mathcal{P}$ labels a point orbit, while $\mathcal{V}$ labels a flat plane. The letters $\mathcal{SP}$ label a two-sphere. The letters $\mathcal{TS}$ label the cotangent bundles to spheres. The letters $\mathcal{U}$ and $\mathcal{H}$ label the hyperboloids orbits of $\SL(2,\bbC)$ and $\SL(2,\bbR)$, with the subscript $0$ labelling the ``degenerate'' hyperboloids, which are cones. Finally, $\mathcal{G}$ labels the paraboloid orbits of the oscillator group. Equation \ref{osc} defines these paraboloids. Interpreting the parameter $\lambda-\lambda_0$ as energy, and the two different components of the vector $\bp$ as position and momentum exhibits equation \ref{osc} to be the equation of a classical harmonic oscillator.
\begin{table}[H]
  \centering
  \caption{Coadjoint Orbits of the Building-Block Groups}
  \label{tab3}
  \rowcolors{2}{blue!10}{white}
\begin{tabular}{c|c|c|c|c} 
 Group & Orbit&Orbit Conditions & Orbit Geometry&Orbit Dimension\\
 \midrule
 $\Aff(1)$ &$\mathcal{P}_{\delta_0}$&$\delta=\delta_0$, $E=0$&pt& $0$  \\ 
    &$\mathcal{V}^\pm$&$E>0$ or $E<0$&$\bbR\times \bbR^\pm$& $2$ \\ 
    \midrule
 $\Euc(2)$  &$\mathcal{P}_{\lambda_0}$&$\lambda=\lambda_0$, $\bp=\boldsymbol{0}$&pt &$0$ \\
    &$\mathcal{TS}^1_{p_0}$&$|\bp|=p_0$&$T^*S^1$ &$2$ \\
\midrule
 $\Euc(3)$  &$\mathcal{P}$&$\boldsymbol{\lambda}=\bp=\boldsymbol{0}$&pt& $0$ \\ 
    &$\mathcal{SP}_{\lambda_0}$&$|\boldsymbol{\lambda}|=\lambda_0$, $\bp=\boldsymbol{0}$&$S^2$& $2$\\ 
    &$\mathcal{TS}^2_{\lambda_0,p_0}$&$|\bp|=p_0$, $\boldsymbol{\lambda}\cdot\bp=\lambda_0p_0$,&$T^*S^2$& $4$ \\
\midrule
$\SU(2)$ &$\mathcal{P}$ &$\boldsymbol{\lambda}=\boldsymbol{0}$&pt& $0$ \\ 
      &$\mathcal{SP}_{\lambda_0}$ &$|\boldsymbol{\lambda}|=\lambda_0$&$S^2$& $2$\\ 
\midrule
 $\SL(2,\bbC)$&$\mathcal{P}$ &$\lambda=0$&pt& $0$ \\
        &$\mathcal{U}_{m}$ &$\det(\lambda)=-m^2$, $m\in \bbC\backslash\{0\}$&$S^2\times \bbR^2$& $4$ \\
        &$\mathcal{U}_0$ &$\det(\lambda)=0$, $\lambda \neq 0$&$\mathbb{RP}^3\times \bbR^+$& $4$ \\
\midrule
 $\SL(2,\bbR)$&$\mathcal{P}$&$\lambda=0$&pt&$0$\\
        &$\mathcal{U}_m$&$\det(\lambda)=-m^2$, $m\neq 0$&$S^1\times \bbR$&$2$\\
        &$\mathcal{H}_m^\pm$&$\det{\lambda}=m^2$, $m\neq 0$, $\lambda_{12}>0$ or $\lambda_{12}<0$&$\bbR^2$&$2$\\
        &$\mathcal{H}^\pm_0$&$\det{\lambda}=0$, $\lambda_{12}>0$ or $\lambda_{12}<0$&$S^1\times \bbR^\pm$&$2$\\
\midrule
 $\N$&$\mathcal{P}_{\bp_0}$&$\bp=\bp_0$, $m=0$&pt&$0$\\
    &$\mathcal{V}_{m_0}$&$m=m_0\neq 0$&$\bbR^2$&$2$\\
\midrule
$\mathcal{D}$&$\mathcal{P}_{\lambda_0}$&$\lambda=\lambda_0$, $\bp=\boldsymbol{0}$, $m=0$&pt&$0$\\
    &$\mathcal{TS}^1_{p_0}$&$|\bp|=p_0$, $m=0$&$T^*S^1$&$2$\\
    &$\mathcal{G}_{\lambda_0,m_0}$&$m=m_0\neq 0$, $\lambda=\lambda_0+\frac{|\bp|^2}{2m_0}\epsilon$&$\bbR^2$&$2$\\
\bottomrule
\end{tabular}
\caption*{The group $\Aff(1)$ is the two-dimensional simply-connected non-abelian group. The euclidean groups in two and three dimensions are denoted $\Euc(2)$ and $\Euc(3)$ respectively. The Heisenberg group is denoted $\N$ and $\mathcal{D}$ denotes the oscillator group.}
\end{table}

In table \ref{tab4a} the UIRs of the building-block groups are listed. Each UIR is labeled with capital letters denoting the carrier space and subscripts denoting the characters that specify the representation. When the carrier space of the representation is infinite dimensional, the inner product is defined by integrating against a (quasi-)invariant integration measure. When the capital letter labeling the UIR corresponds to a calligraphic letter labeling an orbit, it means the orbit and UIR are in correspondence. For example, the UIR $P_{\delta_0}$ of $\Aff(1)$ is the quantisation of the orbit $\mathcal{P}_{\delta_0}$. Sometimes quantisation of an orbit requires quantisation of one of the Casimirs specifying the orbit. An example is in the orbit $\mathcal{G}_{\lambda_0,m_0}$ of the oscillator group, which corresponds to the UIR $G_{n,m_0}$. The real parameter $\lambda_0$ gets quantised to the integer character $n$. The label $P$ denotes the quantisation of a point orbit or of a 2-sphere orbit of $\SU(2)$, which always results in a discrete representation. The label $V$ denotes the quantisation of a plane orbit, which results in a representation carried by square-integrable functions on the line or half-line. The label $TS$ denotes the quantisation of an orbit which is the cotangent bundle to a sphere. The carrier space is the space of sections of certain associated bundles over that same sphere. When $d=2$, there are two such associated bundles, the trivial bundle $E_+$ and the M\"obius bundle $E_-$ over $S^1$. When $d=3$, the integer character $n$ labels the homogeneous line bundles $\mathcal{O}(-n)$ over $S^2\cong \mathbb{CP}^1$. The labels $U$ and $H$ denote the quantisations of the hyperbolic and conical orbits of $\SL(2,\bbC)$ and $\SL(2,\bbR)$. The label $G$ denotes the quantisation of the paraboloid orbit $\mathcal{G}$ of the oscillator group, which is simply the space of standard quantum mechanical wave functions on the real line, with the character $m_0$ playing the part of Planck's constant $\hbar$. Finally, there are the representations labeled $C$. These are the complementary series representations of $\SL(2,\bbC)$ and $\SL(2,\bbR)$. They are the only representations which do not arise out of the quantisation of coadjoint orbits \cite{Gravog}. Table \ref{tab4b} lists the group actions for each of the UIRs, that is, the explicit representations. 
\begin{subtables}
\begin{table}[H]
  \centering
  \caption{UIRs of the Building-Block Groups}
  \label{tab4a}
  \rowcolors{2}{blue!10}{white}
\begin{tabular}{c|c|c|c|c} 
 Group & UIR&Characters & Carrier Space&Integration Measure\\
 \midrule
 $\Aff(1)$ &$P_{\delta_0}$&$\delta_0\in \bbR$&$\bbC$& Discrete\\ 
    &$V^\pm$&&$L^2(\bbR^\pm,\bbC)$& $\frac{dq}{q}$\\ 
\midrule
 $\Euc(2)$  &$P_n$&$n\in \bbZ$&$\bbC$&Discrete\\
    &$TS^\pm_{p_0}$&$p_0\in \bbR^+$&$L^2(S^1,E_\pm)$ &$d\theta$\\
\midrule
 $\Euc(3)$  &$P_n$&$n\in \bbN_0$&$\bbC^{n+1}$& Discrete\\ 
    &$TS^2_{n,p_0}$&$n\in \bbZ$, $p_0\in \bbR^+$&$L^2(S^2,\mathcal{O}(-n))$& $\frac{id\zeta  d\bar{\zeta}}{(1+|\zeta|^2)^2}$\\ 
\midrule
$\SU(2)$ &$P_n$ &$n\in \bbN_0$&$\bbC^{n+1}$& Discrete\\ 
\midrule
 $\SL(2,\bbC)$&$P$ &&$\bbC$& Discrete\\
        &$U_{n,q}$ &$n\in \bbN_0$, $q\in \bbR$&$L^2(\bbC,\bbC)$& $d\zeta  d\bar{\zeta}$\\
        &$C_w$ &$w\in (0,2)$&$L^2(\bbC,\bbC)$& $|\zeta_1-\zeta_2|^{-2+w}d\zeta_1 d\bar{\zeta}_1 d\zeta_2  d\bar{\zeta}_2$\\
\midrule
 $\SL(2,\bbR)$&$P$&&$\bbC$&Discrete\\
         &$U^\pm_q$&$q\in \bbR$&$L^2(\bbR,\bbC)$&$dx$\\
        &$H^{\pm}_n$&$n\in \bbZ$, $n>2$&$L^2(\bbC^+,\bbC)$&$y^{n-2}dx dy$\\
        &$H^{\pm}_0$&&$L^2(\bbC^+,\bbC)$&$dx$\\
         &$C_u$&$u\in (0,1)$&$L^2(\bbR,\bbC)$&$|x-y|^{u-1}dx dy$\\
\midrule
 $\N$&$P_{\bp_0}$&$\bp_0\in \bbR^2$&$\bbC$&Discrete\\
    &$V_{m_0}$&$m_0\in \bbR\backslash\{0\}$&$L^2(\bbR,\bbC)$&$dq$\\
\midrule
$\mathcal{D}$&$P_n$&$n\in \bbZ$&$\bbC$&Discrete\\
    &$TS^\pm_{p_0}$&$p_0\in \bbR^+$&$L^2(S^1,E_\pm)$&$d\theta$\\
    &$G_{n,m_0}$&$n\in \bbZ$, $m_0\in \bbR\backslash\{0\}$&$L^2(\bbR,\bbC)$&$dq$\\
\bottomrule
\end{tabular}
\caption*{Note that $\bbC^+$ refers to the upper half of the complex plane. The trivial and M\"obius bundles over the circle are denoted respectively by $E_+$ and $E_-$, while $\mathcal{O}(-n)$ is the homogeneous bundle of degree $n$ over $\mathbb{CP}^1\cong S^2$. Every UIR corresponds to an orbit except for the complementary series $C_w$ and $C_u$.}
\end{table}
\begin{table}[H]
  \centering
  \caption{UIRs of the Building-Block Groups}
  \label{tab4b}
  \rowcolors{2}{blue!10}{white}
\begin{tabular}{c|w{c}{1cm}|w{c}{9cm}} 
 Group & UIR&Group Action\\
 \midrule
 $\Aff(1)$ &$P_{\delta_0}$&  $\sg(b,a)\cdot \psi=e^{i\delta_0 b}\psi$\\ 
    &$V^\pm$& $[\sg(b,a)\cdot\psi](q)=e^{iaq}\psi(e^{b}q)$\\ 
\midrule
 $\Euc(2)$  &$P_n$&$\sg(R,\bv)\cdot \psi=\nu_n(R)\psi$\\
    &$TS^\pm_{p_0}$& $[\sg(R,\bv)\cdot \psi](\bp)= Q(R,\bp)e^{i\bp\cdot\bv}\psi(R^{-1}\bp)$\\
\midrule
 $\Euc(3)$  &$P_n$& $\sg(R,\bv)\cdot \psi=\rho_n(R)\psi$\\ 
    &$TS^2_{n,p_0}$&$ [\sg(R,\boldsymbol{v})\cdot\psi](\zeta)= (\frac{\bar{a}-b\zeta}{|\bar{a}-b\zeta|})^{-n}e^{i\boldsymbol{p}(\zeta)\cdot\boldsymbol{v}}\psi(\frac{a\zeta+\bar{b}}{-b\zeta+\bar{a}})$\\ 
\midrule
$\SU(2)$ &$P_n$ &$\sg(R)\cdot \psi=\rho_n(R)\psi$\\ 
\midrule
 $\SL(2,\bbC)$&$P$ & Trivial\\
        &$U_{n,q}$ &$[G\cdot F](\zeta)=|-b\zeta+d|^{-2-iq}(\frac{-b\zeta +d}{|-b\zeta+d|})^{-n}F(\frac{a\zeta-c}{-b\zeta+d})$\\
        &$C_w$ &$[G\cdot F](\zeta)=|-b\zeta+d|^{-2-w}F(\frac{a\zeta-c}{-b\zeta+d})$\\
\midrule
 $\SL(2,\bbR)$&$P$&Trivial\\
         &\makecell{$U^+_q$\\$U^-_q$}&\makecell{$[G\cdot F](x)=
            |-bx+d|^{-1-iq}F(\frac{ax-c}{-bx+d})$\\$[G\cdot F](x)=
            \sgn(-bx+d)|-bx+d|^{-1-iq}F(\frac{ax-c}{-bx+d})$}\\
        &\makecell{$H^{+}_n$\\$H^-_n$}&\makecell{ $[G\cdot F](\zeta)=(-b\zeta+d)^{-n}F(\frac{a\zeta-c}{-b\zeta+d)})$\\$[
        G
        \cdot F](\zeta)=(-b \bar{\zeta}+d)^{-n}F(\frac{a\zeta-c}{-b\zeta+d)})$}\\
        &\makecell{$H^{+}_0$\\$H^-_0$}&\makecell{$ [G\cdot F](\zeta)=(-b\zeta+d)^{-1}F(\frac{a\zeta-c}{-b\zeta+d)})$\\$ [G\cdot F](\zeta)=(-b\bar{\zeta}+d)^{-1}F(\frac{a\zeta-c}{-b\zeta+d)})$}\\
         &$C_u$&$ [G\cdot F](x)=|-bx+d|^{-1-u}F(\frac{ax-c}{-bx+d})$\\
\midrule
 $\N$&$P_{\bp_0}$&$ \sg(\bv,\gamma)\cdot \psi = e^{i\bp_0\cdot \bv}\psi$\\
    &$V_{m_0}$&$[\sg(\bv,\gamma)\cdot\psi](q)=e^{i(m_0\gamma+q v_2)}\psi(q+m_0 v_1)$\\
\midrule
$\mathcal{D}$&$P_n$&$\sg(R,\bv,\gamma)\cdot \psi = \nu_n(R)\psi$\\
    &$TS^\pm_{p_0}$&$[\sg(R,\bv,\gamma)\cdot \psi](\bp)=Q(R,\bp)e^{i\bp\cdot \bv}\psi(R^{-1}\bp)$\\
    &$G_{n,m_0}$& $[\sg(R,\boldsymbol{v},\gamma)\cdot\psi](q)= \nu_n(R) e^{i(m_0\gamma+qv_2)}[S(R)\psi](q+m_0 v_1)$\\
\bottomrule
\end{tabular}
\caption*{The function $\nu_n$ for integer $n$ is the degree $n$ representation of the group $\U(1)$. That is, $\nu_n(e^{i\theta}):=e^{in\theta}$. Likewise, $\rho_n$ for positive integer $n$ is the unique $n+1$ dimensional UIR of the group $\SU(2)$. The function $Q(R,\bp)$ is used to account for a possible twist in a bundle over the circle $S^1$. If the bundle is trivial then $Q(R,\bp)=1$. If the bundle is M\"obius, $Q(R,\bp)$ can be $1$ or $-1$. For more detail, see equation \ref{Twist}. The operator $S(R)$ in the group action for the representation $G_{n,m_0}$ of $\mathcal{D}$ is an intertwiner between equivalent representations of the Heisenberg group. For more detail, see appendix \ref{app2}. For the representations of $\SL(2,\bbR)$ and $\SL(2,\bbC)$, $G=\begin{colourpmatrix}a & b\\\hiderowcolors c & d \end{colourpmatrix}$ is a $2\times 2$ matrix with real and complex entries, respectively, such that $ad-bc=1$.}
\end{table}
\end{subtables}
\section{The Indecomposable Groups}

It remains to investigate those Lifshitz groups and central extensions which don't break up into building-blocks. These are the groups $A_3^z$, $A_7$, $C_1^1$ and $C_3$.

\subsection{The group $A_3^z$}
    This Lifshitz group has the additional brackets 
    \begin{equation}
        \begin{split}
            [D,H]&=zH\\
            [D,P_a]&=P_a
        \end{split}
    \end{equation}

where $z$ is a real parameter, characterising the anistropic scaling. When $z=0$, the group can be centrally extended to $C_3$, with the addition of the generator $Z$ such that $[D,H]=Z$. The UIRs and coadjoint orbits of $A_3^0$ can thus be obtained by restricting those of $C_3$ in which $Z$ acts trivially. Therefore, we need only concentrate on the case where $z\neq 0$. We parametrise the group as
\begin{equation}
    \sg(R,b,\bv,t)=e^{\bv\cdot \boldsymbol{P}+tH}Re^{bD}
\end{equation}
with $R \in \Spin(d)$, so that group multiplication is
\begin{equation}
    \sg(R_1,b_1,\bv_1,t_1)\sg(R_2,b_2,\bv_2,t_2)=\sg(R_1 R_2,b_1+b_2,\bv_1+e^{b_1}R\bv_2,t_1+e^{zb_1}t_2)
\end{equation}
We denote an algebra element by $\sX(L,b,\bx,t):=\frac{d}{ds}\sg(R(s),b s, \bx s, t s)|_{s=0}$. A dual element is given by $\sM(\lambda,\delta,\bp,E)$, with the pairing
\begin{equation}
    \langle \sM(\lambda,\delta,\bp,E),\sX(L,b,\bx,t)\rangle := \frac{1}{2}tr(\lambda^T L)+\delta b+ \bp\cdot \bx +Et
\end{equation}

\subsubsection{Coadjoint Orbits}
When $d=2$, the coadjoint action is
\begin{equation}
\begin{split}
    \Ad^*_{\sg(R,b,\bv,t)}\sM(\lambda,\delta,\bp,E)=\sM(\lambda',\delta',\bp',E')
\end{split}
\end{equation}
where
\begin{equation}
    \begin{split}
        \lambda'&=\lambda +e^{-b}\det(R\boldsymbol{p}, \boldsymbol{v})\epsilon\\
        \delta'&= \delta+e^{-b}R\boldsymbol{p}\cdot \boldsymbol{v}+e^{-bz}Ezt\\
        \boldsymbol{p}'&= e^{-b}R\boldsymbol{p}\\
        E'&=e^{-bz}E
    \end{split}
\end{equation}
We thus get the following coadjoint orbits.
\begin{enumerate}
    \item Pointlike orbits given by the equations $\lambda=\lambda_0$, $\delta=\delta_0$, $\bp=\boldsymbol{0}$ and $E=0$ for real number $\delta_0$ and $\mathfrak{so}(2)$ element $\lambda_0$. We label these $\mathcal{P}_{\lambda_0,\delta_0}$.

    \item Choosing a representative where $\bp=\boldsymbol{0}$ and $E\neq 0$ allows us by suitable choice of $t$ and $b$ to arrive at one of two representatives: $\sM(\lambda_0,0,\boldsymbol{0},1)$ or $\sM(\lambda_0,0,\boldsymbol{0},-1)$, depending on the sign of $E$. In each case the stabiliser is $\U(1)\ltimes \bbR^2$ and the orbit is $\bbR\times \bbR^\pm$ characterised by the conditions $\bp=\boldsymbol{0}$, $\lambda=\lambda_0$ and $E>0$ or $E<0$. We label these orbits $\mathcal{V}^\pm_{\lambda_0}$.
    
    \item When $\bp\neq \boldsymbol{0}$, we can choose the representative $\sM(0,0,\boldsymbol{e}_1,E_0)$. This representative is stabilised by $\{\sg(\pm 1, 0,-E_0 zt\boldsymbol{e}_1,t)|t\in \bbR\}\cong \bbZ_2\times \bbR$. The orbit, which we label $\mathcal{F}^1_{E_0}$, has the geometry $T^*S^1\times \bbR^2$ with condition $\bp \neq \boldsymbol{0}$, and is given by the equation
    \begin{equation}
        E=E_0|\bp|^z
    \end{equation}
\end{enumerate}

When $d=3$ we use the $\mathfrak{so}(3)$ adjoint and vector representation equivalence to send $\lambda \in \mathfrak{so}(3)^*$ to $\boldsymbol{\lambda}\in \bbR^3$. The coadjoint action is
\begin{equation}
\begin{split}
\Ad^*_{\sg(R,b,\bv,t)}\sM(\boldsymbol{\lambda},\delta,\bp,E)=\sM(\boldsymbol{\lambda}',\delta',\bp',E')
\end{split}
\end{equation}
where
\begin{equation}
\begin{split}
  \boldsymbol{\lambda}'&=R\boldsymbol{\lambda}+e^{-b}\boldsymbol{v}\times R\boldsymbol{p}\\
  \delta'&=\delta+e^{-b}R\boldsymbol{p}\cdot\boldsymbol{v}+e^{-bz}Ezt\\
      \boldsymbol{p}'&=e^{-b}R\boldsymbol{p}\\
      E'&=e^{-zb}E
    \end{split}
  \end{equation}
We get the following orbits.
\begin{enumerate}
    \item When $\bp=\boldsymbol{0}$ and $E=0$, we can choose a representative $\sM(\lambda_0 \boldsymbol{e}_1,\delta_0,\boldsymbol{0},0)$. When $\lambda_0=0$ we get a pointlike orbit with equations $\lambda=\bp=\boldsymbol{0}$, $E=0$ and $\delta=\delta_0$. We label these $\mathcal{P}_{\delta_0}$. When $\lambda_0>0$, the orbit of this point is the two sphere, given by $|\boldsymbol{\lambda}|=\lambda_0$, $\delta=\delta_0$, $\boldsymbol{p}=\boldsymbol{0}$ and $E=0$. We label these orbits $\mathcal{SP}_{\lambda_0,\delta_0}$

    \item When $\bp=\boldsymbol{0}$ and $E\neq 0$, we can choose a representative of the form $\sM(\lambda_0\boldsymbol{e}_1,0,\boldsymbol{0},\pm 1)$, depending on the sign of $E$. If $\lambda_0=0$, this results in orbits of the form $\bbR \times \bbR^\pm$ with conditions $\boldsymbol{\lambda}=\boldsymbol{0}$, $\bp = \boldsymbol{0}$ and $E>0$ or $E <0$. We call these $\mathcal{V}^\pm$. When $\lambda_0>0$, we get orbits of the form $S^2 \times \bbR\times \bbR^\pm$ constrained by the conditions $|\boldsymbol{\lambda}|=\lambda_0$, $\bp=\boldsymbol{0}$ and $E>0$ or $E<0$ respectively. We call these orbits $\mathcal{SV}^{\pm}_{\lambda_0}$.

    \item When $\bp\neq \boldsymbol{0}$, we can choose a representative of the form $\sM(\lambda_0\boldsymbol{e}_1,0,\boldsymbol{e}_1,E_0)$. The stabiliser of this point is $\{\sg(R,0,-E_0 z t\boldsymbol{e}_1,t)|R\boldsymbol{e}_1=\boldsymbol{e}_1\}\cong \U(1)\times \bbR$. The orbit is given by $\bp\neq 0$, $\boldsymbol{\lambda}\cdot\bp=\lambda_0 |\bp|$ and
    \begin{equation}
        E=E_0|\bp|^z
    \end{equation}
    Changing coordinates from $(\boldsymbol{\lambda},\delta,\bp,E)$ to $(\boldsymbol{k},\delta,\bp,E)$ where $\boldsymbol{k}=\frac{\lambda_0}{|\bp|}-\boldsymbol{\lambda}$ gives us $\boldsymbol{k}\cdot \bp=0$, which exhibits the orbit to be of the form $T^*S^2 \times \bbR^2$. We label these orbits $\mathcal{F}^2_{\lambda_0,E_0}$.
\end{enumerate}
\subsubsection{UIRs}
$A_3^z$ has the Mackey split $(\Spin(d)\times \bbR)\ltimes \bbR^{d+1}$. Since $\bbR^{d+1}$ is abelian, we simply need to investigate the action of $\Spin(d)\times \bbR$ on the dual to $\bbR^{d+1}$, isomorphic to $\bbR^{d+1}$ itself. Let $\tau(\bp,E)$ denote an element of the dual to $\bbR^{d+1}$, with $\bp \in \bbR^d$ and $E \in \bbR$. Let $\sk(R,b)$ denote an element of $\Spin(d)\times \bbR$. Then the action of $\Spin(d)\times \bbR$ on the dual is
\begin{equation}
    \sk(R,b)\cdot \tau(\bp,E)=\tau(e^{-b}R\bp,e^{-zb}E)
\end{equation}

This leads to three classes of $\Spin(d)\times \bbR$ orbit in the dual to $\bbR^{d+1}$:
\begin{enumerate}
    \item The point $\tau(\boldsymbol{0},0)$ is its own orbit. The stabiliser is the whole of $\Spin(d)\times \bbR$.
    
    \item When $\bp= \boldsymbol{0}$ and $E\neq 0$ we can choose a representative $\tau(\boldsymbol{0},1)$ or $\tau(\boldsymbol{0},-1)$. In each case the stabiliser is $\Spin(d)$, and the orbit is either $\bbR^+$ or $\bbR^-$.

    \item When $\bp \neq \boldsymbol{0}$, we can choose a representative of the form $\tau(\boldsymbol{e}_1,E_0)$. The stabiliser of this point is the $\Spin(d-1)$ which fixes $\boldsymbol{e}_1$ and the orbit is $S^{d-1}\times \bbR^+$.
\end{enumerate}
These orbits respectively give rise to the following representations. When $d=2$ we get
\begin{enumerate}
    \item Discrete representations of $A_3^z$ carried by $\bbC$, with the standard inner product. The group action is
    \begin{equation}
        \sg(R,b,\bv,t)\cdot \psi=\nu_n(R)e^{i\delta_0 b}\psi
    \end{equation}
    where $\nu_n(e^{i\theta}):=e^{in\theta}$. The associated characters are $n \in \bbZ$ and $\delta_0 \in \bbR$. We label these representations $P_{n,\delta_0}$.

    \item Representations carried by $L^2(\bbR^+,\bbC)$ and $L^2(\bbR^{-},\bbC)$ with respect to the inner products
    \begin{equation}
        \langle \psi_1,\psi_2\rangle = \int_{\bbR^\pm}\frac{dq}{q}\bar{\psi}_1(q)\psi_2(q)
    \end{equation}

    The group action is 
    \begin{equation}
        [\sg(R,b,\bv,t)\cdot \psi](q)=\nu_n(R)e^{iqt}\psi(e^{bz}q)
    \end{equation}
    The associated character is simply $n \in \bbZ$. We label these representations $V^\pm_{n}$.

    \item Representations carried by sections of the trivial bundle or the twisted Möbius bundle over $S^1$, pulled back over $S^1\times \bbR^+$, which are $L^2$ with respect to the inner product
    \begin{equation}
        \langle \psi_1,\psi_2\rangle = \int_{S^1\times \bbR^+}\frac{d\theta \wedge dq}{q}\bar{\psi}_1(\theta,q)\psi_2(\theta,q)
    \end{equation}
    Let $\boldsymbol{p}=(q\cos \theta, q\sin \theta)^T$. 
    The group action is
     \begin{equation}
        [\sg(R,b,\bv,t)\cdot \psi](\bp)=Q(R,\bp)e^{i\bp\cdot \bv}e^{i E_0 |\bp|^z t} \psi(e^b R^{-1}\bp)
    \end{equation}

    where $Q(R,\bp)$ is $1$ when the bundle is trivial and takes the appropriate value of $1$ or $-1$ when the bundle is M\"obius, according to equation \ref{Twist}. The associated character is $E_0 \in \bbR$. We label these representations $F^\pm_{E_0}$.
\end{enumerate}
When $d=3$ we get
\begin{enumerate}
    \item Discrete representations carried by $\bbC^{n+1}$ with the standard inner product. The group action is
    \begin{equation}
        \sg(R,b,\bv,t)\cdot \psi = \rho_n(R)e^{i\delta_0 b}\psi
    \end{equation}
    where $\rho_n$ is the unique $n+1$ dimensional representation of $\SU(2)$. The associated characters are $n \in \bbN_0$ and $\delta_0 \in \bbR$. We label these representations $P_{n,\delta_0}$.

    \item Representations carried by $L^2(\bbR^+,\bbC)$ and $L^2(\bbR^{-},\bbC)$ with respect to the inner products
    \begin{equation}
        \langle \psi_1,\psi_2\rangle = \int_{\bbR^\pm}\frac{dq}{q}\bar{\psi}_1(q)\psi_2(q)
    \end{equation}
    The group action is
    \begin{equation}
        [\sg(R,b,\bv,t)\cdot \psi](q)=\rho_n(R)e^{iqt}\psi(e^{bz}q)
    \end{equation}

    The associated character is $n \in \bbN_0$. We label these representations $V^\pm_n$.

    \item Representations carried by sections of the bundle $\mathcal{O}(-n)\rightarrow S^2$, pulled back over $S^2\times \bbR^+$, which are $L^2$ with respect to the inner product 
    \begin{equation}
        \langle \psi_1,\psi_2 \rangle= \int_{S^2 \times \bbR^+}\frac{d\zeta \wedge d\bar{\zeta}\wedge dq}{(1+|\zeta|^2)^2 q}\bar{\psi}_1(\zeta,q)\psi_2(\zeta,q)
    \end{equation}

    where $\zeta$ is a homogenous coordinate on the two-sphere and $q$ coordinatises $\bbR^+$. The group action is 
    \begin{equation}
        [\sg(R,b,\boldsymbol{v},t)\cdot\psi](\zeta,q)= (\frac{\bar{a}-b\zeta}{|\bar{a}-b\zeta|})^{-n}e^{i\boldsymbol{p}(\zeta,q)\cdot\boldsymbol{v}}e^{iE_0 q^z t}\psi(\frac{a\zeta+\bar{b}}{-b\zeta+\bar{a}},e^{b}q)
    \end{equation}
    where $\boldsymbol{p}(\zeta,q)$ embeds the point $(\zeta, q)$ on $S^2\times \bbR^+$ into $\bbR^3$. The associated characters are $n \in \bbZ$ and $E_0 \in \bbR$. We label these representations $F^2_{n,E_0}$.
\end{enumerate}
\subsection{The group $A_7$}
This is a two-dimensional Lifshitz group whose brackets, not involving rotations, go as
\begin{equation}
    \begin{split}
        [P_a,P_b]&=\epsilon_{ab}H\\
        [D,P_a]&=P_a\\
        [D,H]&=2H
    \end{split}
\end{equation}
We parametrise the group as $\sg(R,b,\bv,t)=e^{\bv\cdot P}e^{(t-\frac{1}{4}\bv^T B \bv)H}Re^{bD}$, where we recall that 
\begin{equation*}
    B=\begin{pmatrix}
        0&1\\
        1&0
    \end{pmatrix}
\end{equation*}
An algebra element is given as $\sX(L,b,\bx,t)=\frac{d}{ds}\sg(R(s),bs,\bv s,ts)|_{s=0}$. A dual element is given by $\sM(\lambda,\delta,\bp,E)$ with the pairing
\begin{equation}
    \langle \sM(\lambda,\delta,\bp,E),\sX(L,b,\bx,t)\rangle := \frac{1}{2}tr(\lambda^T L)+\delta b+ \bp\cdot \bx +Et
\end{equation}
\subsubsection{Coadjoint Orbits}
The coadjoint actions is calculated to go as
\begin{equation}
\begin{split}
    \Ad_{\sg(R,b,\bv,t)}\sM(\lambda,\delta,\bp,E)=\sM(\lambda',\delta',\bp',E')
\end{split}
\end{equation}
where
\begin{equation}
    \begin{split}
        \lambda'&=\lambda+e^{-b}\det(R\bp,\bv)+\frac{1}{2}e^{-2b}E|\bv|^2 \epsilon\\
        \delta'&=\delta+e^{-b}(R\bp)\cdot \bv +e^{-2b}Et-\frac{1}{2}e^{-2b}E \bv^T B \bv\\
        \bp'&=e^{-b}R\bp+e^{-2b}E\epsilon \bv\\
        E'&=e^{-2b}E
    \end{split}
\end{equation}
where we recall 
\begin{equation}
    \epsilon=\begin{pmatrix}
        0&1\\
        -1&0
    \end{pmatrix}
\end{equation}
We get the following orbits
\begin{enumerate}
    \item Pointlike orbits given by the equations $\bp=\boldsymbol{0}$, $E=0$, $\lambda=\lambda_0$ and $\delta=\delta_0$. We label these $\mathcal{P}_{\lambda_0,\delta_0}$

    \item When $E=0$ and $\bp\neq \boldsymbol{0}$, we can choose a representative $\sM(0,0,\boldsymbol{e}_1,0)$. The orbit this gives rise to has the geometry $T^*S^1\times \bbR^2$ and is given by the conditions $\bp \neq \boldsymbol{0}$ and $E=0$. We label this orbit $\mathcal{TSV}^1$.

    \item When $E \neq 0$, we can choose representatives $\sM(\lambda_0,0,\boldsymbol{0},1)$ or $\sM(\lambda_0,0,\boldsymbol{0},-1)$. The orbits these give rise to, which we label $\mathcal{G}^\pm_{\lambda_0}$, are of the form $\bbR^3\times \bbR^\pm$ respectively, given by the respective conditions $E>0$ or $E<0$, and the equation
    \begin{equation}
        \lambda=\lambda_0+\frac{1}{2E}|\bp|^2 \epsilon
    \end{equation}
\end{enumerate}
\subsubsection{UIRs}
$A_7$ has the Mackey split $(\U(1)\times \bbR)\ltimes \N$. We need to consider $\U(1)\times \bbR$ orbits in the unitary dual of the Heisenberg group. Recall that $\hat{\N}$ consists of a plane of discrete representations where the central element acts trivially, and a punctured line of continuous representations. Let $\sk(R,b)\in \U(1)\times \bbR$. Then, for a representation $\chi \in \hat{\N}$ and an element $\st(\bv,t)\in \N$, we have that
\begin{equation}
    [\sk\cdot \chi](\st)=\chi(\sk^{-1}\cdot \st)
\end{equation}
When we perform the calculations we find that
\begin{equation}
    \sk(R,b)^{-1}\cdot \st(\bv,t) = \st(e^{-b}R^{-1}\bv,e^{-2b}t)
\end{equation}
The representations change as follows. The representations on the plane are specified by a character $\bp \in \bbR^2$. This character becomes $e^{-b}R\bp$. The continuous representations are specified by $E \in \bbR\backslash\{0\}$. This character becomes $e^{-2b}E$. We thus get the following orbits in $\hat{\N}$:
\begin{enumerate}
    \item A point orbit at the origin.

    \item A cylindrical orbit consisting  of every point in the plane besides the origin.

    \item Two ray-orbits, partitioning the vertical line into positive and negative halves.
\end{enumerate}
These give rise to the following representations of $A_7$.
\begin{enumerate}
    \item A discrete representation carried by $\bbC$ with the standard inner product. The group action is
    \begin{equation}
        \sg(R,b,\bv,t)\cdot \psi = \nu_n(R)e^{i\delta_0 b}\psi
    \end{equation}
    where $\nu_n(e^{i\theta})=e^{in\theta}$. The associated characters are $n \in \bbZ$ and $\delta_0\in \bbR$. We label these representations $P_{n,\delta_0}$.

    \item Representations carried by sections of either the trivial or M\"obius bundles over $S^1$, pulled back over the cylinder $S^1\times \bbR^+$, which are $L^2$ with respect to the inner product
    \begin{equation}
        \langle \psi_1,\psi_2\rangle = \int_{S^1\times \bbR^+}\frac{d\theta \wedge dq}{q}\bar{\psi}_1(\theta,q)\psi_2(\theta,q)
    \end{equation}
    Let $\boldsymbol{p}=(q\cos \theta, q\sin \theta)^T$. 
    The group action is
    \begin{equation}
        [\sg(R,b,\bv,t)\cdot \psi](\bp)=Q(R,\bp)e^{i\bp\cdot \bv}\psi(e^b R^{-1}\bp)
    \end{equation}
    where $Q(R,\bp)$ accounts for the twist if the bundle is M\"obius. These representations we label $TSV^\pm$.

    \item Representations carried by sections of Hilbert bundles over $\bbR^+$ and $\bbR^-$. Each fibre of these Hilbert bundles is a space of functions $\bbR\rightarrow \bbC$. We can more cleanly think of the representation spaces as the spaces of functions $\psi_\pm:\bbR\times \bbR^\pm \rightarrow \bbC$, which are $L^2$ with respect to
    \begin{equation}
        \langle \psi_1,\psi_2\rangle= \int_{\bbR\times \bbR^\pm}\frac{dx\wedge dy}{y}\bar{\psi}_1(x,y)\psi_2(x,y)
    \end{equation}

    The group action is 
    \begin{equation}
        [\sg(R,b,\bv,t)\cdot \psi_\pm](x,y)=\nu_n(R)e^{iyt+i|y|^\frac{1}{2}xv_2}[S(R)\psi_\pm] (x\pm |y|^\frac{1}{2}v_1,e^{2b}y)
    \end{equation}
    where $S(R)$ accounts for the fact that $\U(1)$ sends the continuous representations to equivalent but different representations in $\hat{N}$. It is an integral in the first coordinate $x$, as in equation \ref{Mehler}. The associated character is $n\in \bbZ$. We label these representations $G^\pm_n$.
\end{enumerate}
\subsection{The group $C_1^1$}
$C_1^1$ is a central extension of $A_1$ in $2$ dimensions by a generator $Z$. The brackets not involving the rotations are
\begin{equation}
    \begin{split}
        [D,H]&=Z\\
        [P_a,P_b]&=\epsilon_{ab}Z
    \end{split}
\end{equation}
We parametrise the group as
\begin{equation}
    \sg(R,b,\bv,t,z)=e^{\bv\cdot\boldsymbol{P}+tH}e^{(z-\frac{1}{4}\bv^T B \bv)Z}Re^{bD}
\end{equation}
An algebra element is given as $\sX(L,b,\bx,t,u)=\frac{d}{ds}\sg(R(s),bs,\bx s,ts,us)|_{s=0}$. A dual element is $\sM(\lambda,\delta,\bp,E,m)$ and the pairing is
\begin{equation}
    \langle \sM(\lambda,\delta,\bp,E,m),\sX(L,b,\bx,t,u)\rangle := \frac{1}{2}tr(\lambda^T L)+\delta b+ \bp\cdot \bx +Et+mu
\end{equation}
\subsubsection{Coadjoint Orbits}
The coadjoint action is 
\begin{equation}
\begin{split}
\Ad^*_{\sg(R,b,\bv,t,z)}\sM(\lambda,\delta,\bp,E,m)=\sM(\lambda',\delta',\bp',E',m')
\end{split}
\end{equation}
where
\begin{equation}
    \begin{split}
        \lambda'&=\lambda+\det(R\bp,\bv)\epsilon+\frac{1}{2}m|\bv|^2\epsilon\\
        \delta'&=\delta+mt\\
        \bp'&=R\bp+m\epsilon \bv\\
        E'&=E-mb\\
        m'&=m
    \end{split}
\end{equation}
We get the following orbits
\begin{enumerate}
    \item When $m$ and $\bp$ are both zero we get point orbits given by the equations $\lambda=\lambda_0$, $\delta=\delta_0$, $\bp=\boldsymbol{0}$, $E=E_0$ and $m=0$. We label these $\mathcal{P}_{\lambda_0,\delta_0,E_0}$.

    \item When $\bp\neq \boldsymbol{0}$ and $m=0$, we can choose an orbit representative $\sM(0,\delta_0,p_0 \boldsymbol{e}_1,E_0,0)$. The orbit of this representative has the geometry $T^*S^1$ with equations $\delta=\delta_0$, $E=E_0$, $m=0$ and $|\bp|=p_0$. We label these $\mathcal{TS}^1_{\delta_0,p_0,E_0}$.

    \item When $m \neq 0$, we can choose a representative $\sM(\lambda_0,0,\boldsymbol{0},0,m_0)$. The orbit of this point has the geometry $\bbR^4$, given by the equations $m=m_0$ and
    \begin{equation}
        \lambda=\lambda_0+\frac{1}{2m}|\bp|^2 \epsilon
    \end{equation}
    The name we give to these orbits is $\mathcal{G}_{\lambda_0,m_0}$.
\end{enumerate}

\subsubsection{UIRs}
$C_1^1$ has the Mackey split $(\U(1)\times \bbR)\ltimes (\N \times \bbR)$. The unitary dual of $\N\times \bbR$ is simply $\hat{\N}\times \bbR$ and we need to look at $\U(1)\times \bbR$ orbits in this space. Let $\sk(R,b)$ denote an element in $\U(1)\times \bbR$ and let $\st(\bv,t,z)$ denote an element in $\N\times \bbR$. Let $\chi$ denote a representation of $\N\times \bbR$. Then we have
\begin{equation}
    [\sk\cdot \chi](\st)=\chi(\sk^{-1}\cdot \st)
\end{equation}
Doing the calculation we find
\begin{equation}
    \sk(R,b)^{-1}\cdot \st(\bv,t,z)=\st(R^{-1}\bv,t,z-\frac{1}{4}\bv^T B \bv+\frac{1}{4}(R^{-1}\bv)^T B (R^{-1}\bv))
\end{equation}
The discrete representations of $\N\times \bbR$ are indexed by a pair $(\bp,E)$ with $\bp\in \bbR^2$ and $E \in \bbR$. Under the action of $\U(1)\times \bbR$ this becomes $(R\bp,E)$. The continuous representations are indexed by a pair $(m,E)$. Under the $\U(1)\times \bbR$ action this becomes $(m,E-m b)$. We thus get the following orbits in $\hat{\N}\times \bbR$
\begin{enumerate}
    \item Point orbits, given by $(\boldsymbol{0},E_0)$.

    \item $S^1$ orbits in the plane given by $\{(\bp,E_0)||\bp|=p_0\}$.

    \item Orbits diffeomorphic to $\bbR$, given by $\{(m_0,-m_0 b)|b\in \bbR\}$ for fixed $m_0$.
\end{enumerate}
These give rise to the following representations
\begin{enumerate}
    \item Discrete representations carried by $\bbC$ with the standard inner product. The group action is
    \begin{equation}
        \sg(R,b,\bv,t,z)\cdot \psi = \nu_n(R)e^{i\delta_0 b}e^{iE_0t}\psi
    \end{equation}
    The associated characters are $n \in \bbZ$, $\delta_0 \in \bbR$ and $E_0 \in \bbR$. We label these $P_{n,\delta_0, E_0}$.

    \item Representations carried by sections of the trivial or M\"obius bundles over $S^1_{p_0}$, the sphere with radius $p_0$, which are $L^2$ with respect to the inner product
    \begin{equation}
        \langle \psi_1,\psi_2\rangle = \int_{S^1}\bar{\psi}_1(\theta)\psi_2(\theta)d\theta
    \end{equation}
    Let $\bp=(p_0\cos \theta, p_0\sin \theta)^T$. The group action is
    \begin{equation}
        [\sg(R,b,\bv,t,z)\cdot \psi](\bp)=Q(R,\bp)e^{i\delta_0 b}e^{i\bp\cdot \bv}e^{iE_0 t}\psi(R^{-1}\bp)
    \end{equation}
    where $Q(R,\bp)$ accounts for the possible twist in the bundle. The associated characters are $\delta_0\in \bbR$, $p_0 \in \bbR^+$ and $E_0\in \bbR$. We label these representations $TS^\pm_{\delta_0,p_0,E_0}$.

    \item Representations which are sections of a Hilbert bundle over $\bbR$. Each fibre is a space of functions $\bbR\rightarrow \bbC$. The correct way of thinking of the space as is functions $\psi: \bbR^2\rightarrow \bbC$, which are $L^2$ with respect to the inner product
    \begin{equation}
        \langle \psi_1,\psi_2\rangle=\int_{\bbR^2} dx\wedge dy \bar{\psi}_1(x,y)\psi_2(x,y)
    \end{equation}
    The group action is
    \begin{equation}
        [\sg(R,b,\bv,t,z)\cdot \psi](x,y)=\nu_n(R)e^{i(m_0 z+yt+xv_2)}[S(R)\psi](x+m_0 v_1,y+m_0 b)
    \end{equation}
    where $S(R)$ is an integral in the first coordinate $x$ as in equation \ref{Mehler}. The associated characters are $n \in \bbZ$ and $m_0 \in \bbR\backslash\{0\}$. These representations we label $G_{n,m_0}$.
\end{enumerate}
\subsection{The  group $C_3$}
$C_3$ is a central extension of $A_3^0$. There is an extra generator, $Z$, and the non-zero brackets which don't involve the rotation generators are
\begin{equation}
    \begin{split}
        [D,P_a]&=P_a\\
        [D,H]&=Z
    \end{split}
\end{equation}

We parametrise the group as $\sg(R,b,\bv,t,z)=e^{\bv \cdot \boldsymbol{P}+tH}Re^{bD}e^{zZ}$ with $R \in \Spin(d)$. 
We denote an adjoint element by $\sX(L,b,\bx,t,u)=\frac{d}{ds}\sg(R(s),bs,\bx s, ts, us)|_{s=0}$ and a coadjoint orbit by $\sM(\lambda,\delta,\bp,E,m)$, where the pairing is
\begin{equation}
    \langle \sM(\lambda,\delta,\bp,E,m),\sX(L,b,\bx,t,u)\rangle := \frac{1}{2}tr(\lambda^T L)+\delta b+ \bp\cdot \bx +Et+mu
\end{equation}
\subsubsection{Coadjoint Orbits}
When $d=2$ the coadjoint action is
\begin{equation}
\begin{split}
\Ad^*_{\sg(R,b,\bv,t,z)}\sM(\lambda,\delta,\bp,E,m)=\sM(\lambda',\delta',\bp',E',m')
\end{split}
\end{equation}
where
\begin{equation}
    \begin{split}
        \lambda'&=\lambda +e^{-b}\det(R\boldsymbol{p}, \boldsymbol{v})\epsilon\\
        \delta'&= \delta+e^{-b}R\boldsymbol{p}\cdot \boldsymbol{v}+mt\\
        \boldsymbol{p}'&= e^{-b}R\boldsymbol{p}\\
        E'&=E-mb\\
        m'&=m
    \end{split}
\end{equation}

\begin{enumerate}
    \item We get pointlike orbits given by the equations $\lambda =\lambda_0$, $\delta =\delta_0$, $\bp=\boldsymbol{0}$, $E=E_0$ and $m=0$. We label these $\mathcal{P}_{\lambda_0,\delta_0,E_0}$.

    \item When $m \neq 0$ and $\bp=\boldsymbol{0}$ we can choose a representative of the form $\sM(\lambda_0,0,\boldsymbol{0},0,m_0)$. The orbit of this point is diffeomorphic to $\bbR^2$ and is given by the equations $\lambda =\lambda_0$, $\bp =\boldsymbol{0}$ and $m=m_0$. We label these orbits $\mathcal{V}_{\lambda_0,m_0}$.

    \item When $\bp\neq \boldsymbol{0}$ we can choose a representative of the form $\sM(0,0,\boldsymbol{e}_1,E_0,m_0)$. The stabiliser of this point is $\{\sg(\pm 1, 0,-m_0 t \boldsymbol{e}_1,t,z)|t,z\in \bbR\}\cong \bbZ_2 \times \bbR^2$. The orbits have the geometry $T^*S^1 \times \bbR^2$, given by the relations $\bp \neq \boldsymbol{0}$, $m=m_0$ and
    \begin{equation}
        E=E_0+m_0 \log|\bp|
    \end{equation}
    We label these orbits $\mathcal{Y}^1_{E_0,m_0}$.
\end{enumerate}
When $d=3$ we again leverage the properties of the representations of $\mathfrak{so}(3)$ to send $\lambda \in \mathfrak{so}(3)^*$ to $\boldsymbol{\lambda}\in \bbR^3$. The coadjoint action is
\begin{equation}
\begin{split}
\Ad^*_{\sg(R,b,\bv,t,z)}\sM(\boldsymbol{\lambda},\delta,\bp,E,m)=\sM(\boldsymbol{\lambda}',\delta',\bp',E',m')
\end{split}
\end{equation}
where
\begin{equation}
    \begin{split}
        \boldsymbol{\lambda}'&=R\boldsymbol{\lambda}+e^{-b}\bv \times R\bp\\
        \delta'&= \delta+e^{-b}R\boldsymbol{p}\cdot \boldsymbol{v}+mt\\
        \boldsymbol{p}'&= e^{-b}R\boldsymbol{p}\\
        E'&=E-mb\\
        m'&=m
    \end{split}
\end{equation}

\begin{enumerate}
    \item When $\bp =\boldsymbol{0}$ and $m=0$, we can choose a representative $\sM(\lambda_0 \boldsymbol{e}_1,\delta_0,\boldsymbol{0},E_0,0)$. When $\lambda_0=0$ this results in pointlike orbits given by the equations $\boldsymbol{\lambda}=\bp=\boldsymbol{0}$, $\delta =\delta_0$, $E=E_0$ and $m=0$. These we call $\mathcal{P}_{\delta_0,E_0}$. When $\lambda_0 >0$, we get $S^2$ orbits given by the equations $|\boldsymbol{\lambda}|=\lambda_0$, $\bp=\boldsymbol{0}$, $\delta =\delta_0$, $E=E_0$ and $m=0$. We call these $\mathcal{SP}_{\lambda_0,\delta_0,E_0}$.

    \item When $\bp =\boldsymbol{0}$ and $m\neq0$ we can choose a representative $\sM(\lambda_0 \boldsymbol{e}_1,0,\boldsymbol{0},0,m_0)$. When $\lambda_0=0$, this point has $\bbR^2$ orbits, given by the equations $\boldsymbol{\lambda}=\bp=\boldsymbol{0}$ and $m=m_0$. We call these $\mathcal{V}_{m_0}$. When $\lambda_0 >0$, we get $S^2\times \bbR^2$ orbits given by the equations $|\boldsymbol{\lambda}|=\lambda_0$, $\bp=\boldsymbol{0}$ and $m=m_0$. We call these $\mathcal{SV}_{\lambda_0,m_0}$

    \item When $\bp\neq \boldsymbol{0}$, we can choose a representative $\sM(\lambda_0 \boldsymbol{e}_1, 0, \boldsymbol{e}_1, E_0, m_0)$. The orbit of this point is $T^*S^2 \times \bbR^2$, given by the conditions $\bp \neq 0$, $\boldsymbol{\lambda}\cdot \bp = \lambda_0|\bp|$, $m = m_0$ and 
    \begin{equation}
        E=E_0+m_0\log|\bp|
    \end{equation}
    We label these $\mathcal{Y}^2_{\lambda_0,E_0,m_0}$.
\end{enumerate}

\subsubsection{UIRs}
$C_3$ has the Mackey split $(\Spin(d)\times \bbR)\ltimes \bbR^{d+2}$. We consider the action of $\Spin(d)\times \bbR$ on the dual to $\bbR^{d+2}$. Let $\sk(R,b)$ denote an element of $\Spin(d)\times \bbR$ while $\tau(\bp,t,m)$ denotes an element of the dual to $\bbR^{d+2}$. We have
\begin{equation}
    \sk(R,b)\cdot \tau(\bp,t,m)=\tau(e^{-b}R\bp,E-mb,m)
\end{equation}
We get the following orbits
\begin{enumerate}
    \item Pointlike orbits of the form $\tau(\boldsymbol{0},E_0,0)$. The stabiliser is the whole group $\Spin(d)\times \bbR$.

    \item When $\bp=\boldsymbol{0}$ and $m\neq 0$ we can choose a representative of the form $\tau(\boldsymbol{0},0,m_0)$. The stabiliser of this point is $\Spin(d)$ and the orbit is diffeomorphic to $\bbR$.

    \item When $\bp \neq \boldsymbol{0}$, we can choose a representative of the form $\tau(\boldsymbol{e}_1,E_0,m_0)$. The stabiliser of this point is the $\Spin(d-1)$ which stabilises $\boldsymbol{e}_1$. The orbit is $S^{d-1} \times \bbR^+$.
\end{enumerate}

These correspond to the following representations. For $d=2$ we get
\begin{enumerate}
    \item Representations carried by $\bbC$ with the standard inner product. The group action is 
    \begin{equation}
        \sg(R,b,\bv,t,z)\cdot \psi=\nu_n (R)e^{i\delta_0 b}e^{iE_0 t}\psi
    \end{equation}
    The associated characters are $n \in \bbZ$, $\delta_0\in \bbR$ and $E_0 \in \bbR$. We label these representations $P_{n,\delta_0,E_0}$.

    \item Representations carried by $L^2(R,\bbC)$ with respect to the standard inner product given by integrating over $\bbR$. The group action is
    \begin{equation}
        [\sg(R,b,\bv,t,z)\cdot \psi](q)=\nu_n(R)e^{im_0 z+ iqt}\psi(q+bm_0)
    \end{equation}
    The associated characters are $n\in \bbZ$ and $m_0 \in \bbR\backslash\{0\}$. These representations we call $V_{n,m_0}$.

    \item Representations carried by sections of either the trivial or M\"obius bundles over $S^1$, pulled back over $S^1\times \bbR$, which are $L^2$ with respect to
     \begin{equation}
        \langle \psi_1,\psi_2\rangle = \int_{S^1\times \bbR^+}\frac{d\theta \wedge dq}{q}\bar{\psi}_1(\theta,q)\psi_2(\theta,q)
    \end{equation}
    Let $\bp=(q\cos \theta, q \sin \theta)^T$. The group action is 
    \begin{equation}
        [\sg(R,b,\bv,t,z)\cdot \psi](\bp)=Q(R,\bp)e^{i\bp\cdot \bv}e^{i(E_0+m_0\log|\bp|) t}e^{im_0 z}\psi(e^b R^{-1}\bp)
    \end{equation}
    where $Q(R,\bp)$ accounts for the possible twisting in the bundle. The associated characters are $E_0 \in \bbR$ and $m_0 \in \bbR$. We call these representations $Y^\pm_{E_0,m_0}$.
\end{enumerate}

For $d=3$ we get
\begin{enumerate}
    \item Representations carried by $\bbC^{n+1}$ with the standard inner product. The group action is
    \begin{equation}
        \sg(R,b,\bv,t,z)\cdot \psi=\rho_n(R)e^{i\delta_0 b}e^{iE_0 t}\psi
    \end{equation}
    with associated characters $n \in \bbN_0$, $\delta_0 \in \bbR$ and $E_0 \in \bbR$. We call these representations $P_{n,\delta_0,E_0}$.

    \item Representations carried by $L^2(\bbR,\bbC)$ with respect to the standard inner product. The group action is
    \begin{equation}
        [\sg(R,b,\bv,t,z)\cdot\psi](q)=\rho_n(R)e^{im_0 z+ iqt}\psi(q+bm_0)
    \end{equation}
    The associated characters are $n \in \bbN_0$ and $m_0\in \bbR\backslash\{0\}$. These representations we label $V_{n,m_0}$.

    \item Representations carried by sections of the bundle $\mathcal{O}(-n)\rightarrow S^2$, pulled back over $S^2\times \bbR^+$, which are $L^2$ with respect to the inner product
     \begin{equation}
        \langle \psi_1,\psi_2 \rangle= \int_{S^2 \times \bbR^+}\frac{d\zeta \wedge d\bar{\zeta}\wedge dq}{(1+|\zeta|^2)^2 q}\bar{\psi}_1(\zeta,q)\psi_2(\zeta,q)
    \end{equation}
    The group action is
    \begin{equation}
        [\sg(R,b,\boldsymbol{v},t,z)\cdot\psi](\zeta,q)= (\frac{\bar{a}-b\zeta}{|\bar{a}-b\zeta|})^{-n}e^{i\boldsymbol{p}(\zeta,q)\cdot\boldsymbol{v}}e^{i(E_0 +m_0\log|\bp(\zeta,q)|) t}e^{im_0 z}\psi(\frac{a\zeta+\bar{b}}{-b\zeta+\bar{a}},e^{b}q)
    \end{equation}
    The associated characters are $n \in \bbZ$, $E_0 \in \bbR$ and $m_0\in \bbR$. These representations we call $Y^2_{n.E_0,m_0}$.
\end{enumerate}

\subsection{Summary of Results}\label{results2} The coadjoint orbits of the indecomposable groups are listed below in table \ref{tab5}. The calligraphic capital letters denote the geometry of each orbit and how it is embedded in the coalgebra. The subscripts denote the Casimirs on the orbit. The orbits labeled $\mathcal{P}$ are point orbits, while those labeled $\mathcal{V}$ are planes. The orbits labeled $\mathcal{SP}$ are two-spheres and those labeled $\mathcal{SV}$ are the product of a two-sphere with a plane. Those labeled $\mathcal{TS}$ are the cotangent bundles of spheres and those labeled $\mathcal{TSV}$ are the product of contangent bundles of spheres with planes. These orbits are ``non-generic'' in the sense that they are confined to specific lower dimensional subspaces of the coalgebra where at least one of the coordinates is zero. Each group also has ``generic orbits'' in the sense that if you pick a random point in the coalgebra it is likely to lie in one of these orbits. The group $A_3^z$ has the generic orbit $\mathcal{F}$. This orbit is specified by the equation
\begin{equation}\label{disp}
    E=E_0|\bp|^z
\end{equation}
This can be viewed as the dispersion relation for a Lifshitz system with symmetry group $A_3^z$. There is an initial energy scale, $E_0$, and the parameter $z$ tells us how energy scales with momentum. The groups $A_7$ and $C_1^1$ both have generic paraboloid orbits $\mathcal{G}$ similar to those of the oscillator group. The equations specifying these orbits are respectively
\begin{equation}\label{spin1}
    \lambda=\lambda_0+\frac{|\bp|^2}{2E}\epsilon
\end{equation}
and
\begin{equation}\label{spin2}
    \lambda=\lambda_0+\frac{|\bp|^2}{2m_0}\epsilon
\end{equation}
We can view equation \ref{spin1} as a relationship between spin ($\lambda$), energy ($E$) and momentum ($\bp$) in a classical Lifshitz system with $A_7$ symmetry. Likewise we can view \ref{spin2} as a relationship between spin and momentum in a system with $A_1$ symmetry.
Finally, the group $C_3$ has generic orbits $\mathcal{Y}$ specified by 
\begin{equation}
    E=E_0+m_0\log|\bp|
\end{equation}
This can also be viewed as a dispersion relation for a classical system with $A_3^0$ symmetry, with energy growing logarithmically with momentum.
\begin{table}[H]
  \centering
  \caption{Coadjoint Orbits of the Indecomposable Groups}
  \label{tab5}
  \rowcolors{2}{blue!10}{white}
  \resizebox{\linewidth}{!}{
\begin{tabular}{c|c|c|c|c} 
 Group & Orbit&Orbit Conditions & Orbit Geometry&Orbit Dimension\\
 \midrule
 $A_3^z$, $d=2$ &$\mathcal{P}_{\lambda_0,\delta_0}$&$\lambda=\lambda_0$, $\delta=\delta_0$, $\bp=\boldsymbol{0}$, $E=0$&pt& $0$ \\ 
    &$\mathcal{V}^\pm_{\lambda_0}$&$\lambda=\lambda_0$, $\bp=\boldsymbol{0}$, $E>0$ or $E<0$&$\bbR\times \bbR^\pm$& $2$ \\ 
    &$\mathcal{F}^1_{E_0}$&$\bp\neq \boldsymbol{0}$, $E=E_0|\bp|^z$&$T^*S^1\times\bbR^2$& $4$  \\ 
\midrule
 $A_3^z$, $d=3$  &$\mathcal{P}_{\delta_0}$&$\boldsymbol{\lambda}=\bp=\boldsymbol{0}$, $\delta=\delta_0$, $E=0$&pt &$0$ \\
    &$\mathcal{SP}_{\lambda_0,\delta_0}$& $|\boldsymbol{\lambda}|=\lambda_0$, $\delta=\delta_0$, $\bp=\boldsymbol{0}$, $E=0$&$S^2$ &$2$\\
    &$\mathcal{V}^\pm$& $\boldsymbol{\lambda}=\bp=\boldsymbol{0}$, $E>0$ or $E<0$&$\bbR\times \bbR^\pm$ &$2$\\
    &$\mathcal{SV}^\pm_{\lambda_0}$& $|\boldsymbol{\lambda}|=\lambda_0$, $\bp=\boldsymbol{0}$, $E>0$ or $E<0$&$S^2\times \bbR\times \bbR^\pm$ & $4$\\
    &$\mathcal{F}^2_{\lambda_0,E_0}$&$\bp\neq \boldsymbol{0}$, $\boldsymbol{\lambda}\cdot\bp=\lambda_0|\bp|$, $E=E_0|\bp|^z$&$T^*S^2\times \bbR^2$ &$6$\\
\midrule
 $A_7$  &$\mathcal{P}_{\lambda_0,\delta_0}$&$\lambda=\lambda_0$, $\delta=\delta_0$, $\bp=\boldsymbol{0}$, $E=0$&pt& $0$\\ 
    &$\mathcal{TSV}^1$&$\bp\neq \boldsymbol{0}$, $E=0$&$T^*S^1\times \bbR^2$& $4$\\ 
    &$\mathcal{G}^\pm_{\lambda_0}$&$E>0$ or $E<0$, $\lambda=\lambda_0+\frac{|\bp|^2}{2E}\epsilon$&$\bbR^3\times \bbR^\pm$& $4$ \\ 
\midrule
 $C_1^1$&$\mathcal{P}_{\lambda_0,\delta_0,E_0}$ &$\lambda=\lambda_0$, $\delta=\delta_0$, $\bp=\boldsymbol{0}$, $E=E_0$, $m=0$&pt& $0$\\
    &$\mathcal{TS}^1_{\delta_0,p_0,E_0}$ &$\delta=\delta_0$, $|\bp|=p_0$, $E=E_0$, $m=0$&$T^*S^1$& $2$\\
    &$\mathcal{G}_{\lambda_0,m_0}$ &$m=m_0\neq 0$, $\lambda=\lambda_0+\frac{|\bp|^2}{2m}\epsilon$&$\bbR^4$& $4$\\
\midrule
 $C_3$, $d=2$&$\mathcal{P}_{\lambda_0,\delta_0,E_0}$&$\lambda=\lambda_0$, $\delta=\delta_0$, $\bp=\boldsymbol{0}$, $E=E_0$, $m=0$&pt&$0$\\
    &$\mathcal{V}_{\lambda_0,m_0}$&$\lambda=\lambda_0$, $\bp=\boldsymbol{0}$, $m=m_0\neq 0$&$\bbR^2$&$2$\\
    &$\mathcal{Y}^1_{E_0,m_0}$&$\bp\neq \boldsymbol{0}$, $m=m_0$, $E=E_0+m_0 \log|\bp|$&$T^*S^1\times \bbR^2$&$4$\\
\midrule
$C_3$, $d=3$&$\mathcal{P}_{\delta_0,E_0}$&$\boldsymbol{\lambda}=\boldsymbol{0}$, $\delta=\delta_0$, $\bp=\boldsymbol{0}$, $E=E_0$, $m=0$&pt&$0$\\
    &$\mathcal{SP}_{\lambda_0,\delta_0,E_0}$&$|\boldsymbol{\lambda}|=\lambda_0$, $\delta=\delta_0$, $\bp=\boldsymbol{0}$, $E=E_0$, $m=0$&$S^2$&$2$\\
    &$\mathcal{V}_{m_0}$&$\boldsymbol{\lambda}=\bp=\boldsymbol{0}$, $m=m_0\neq 0$&$\bbR^2$&$2$\\
    &$\mathcal{SV}_{\lambda_0,m_0}$&$|\boldsymbol{\lambda}|=\lambda_0$, $\bp=\boldsymbol{0}$, $m=m_0\neq 0$&$S^2\times\bbR^2$&$4$\\
    &$\mathcal{Y}^2_{\lambda_0,E_0,m_0}$&$\bp\neq\boldsymbol{0}$, $m=m_0$, $\boldsymbol{\lambda}\cdot \bp=\lambda_0|\bp|$, $E=E_0+m_0\log|\bp|$&$T^*S^2\times\bbR^2$&$6$\\
    \bottomrule
\end{tabular}
}
\caption*{In $A_3^z$, $d=3$ and $C_3$, $d=3$, the orbits labelled $\mathcal{SP}$ and $\mathcal{SV}$ are the products of an $S^2$ with the orbits labelled $\mathcal{P}$ and $\mathcal{V}$ respectively. When the radius of this $S^2$ goes to zero, we get the cases $\mathcal{P}$ and $\mathcal{V}$.}
\end{table}
The UIRs of the indecomposable groups are listed in table $\ref{tab6a}$. The capital letters labeling each representation refer to the representation space while the subscripts denote the characters specifying the representation. When capital letters denoting a UIR correspond to the caligraphic capital letters denoting a coadjoint orbit, it means the two are in correspondence. For example, the representation $F^2_{n,E_0}$ of $A_3^z$ with $d=3$ is obtained by quantising the orbit $\mathcal{F}^2_{\lambda_0,E_0}$. In this case, the real parameter $\lambda_0$ is quantised to get the integer character $n$. The UIRs labeled $P$ are the quantisations of the point-like and spherical orbits. These have discrete carrier spaces. The UIRs labeled $V$ arise from quantising plane orbits or orbits which are spheres cross planes. They are carried by spaces of square integrable functions on the real line or half-line. The UIRs labeled $TSV^\pm$ are the quantisation of the orbit which is the cotangent bundle of the two-sphere cross a plane. It is carried by sections of the trivial or M\"obius bundle, pulled back over the cylinder $S^1\times \bbR^+$. The UIRS labeled $TS^\pm$ are the quantisation of the cotangent bundle over the circle. They are carried by sections of the trivial or M\"obius bundles over the circle. Finally, there are the quantisations of the generic orbits. The representations labeled $F$ and $Y$ are the quantisations of the orbits labeled $\mathcal{F}$ and $\mathcal{Y}$ respectively. They are both given by sections of associated bundles over cylinders. When $d=2$, the carrier space is the space of sections of either the trivial or M\"obius bundle over the circle, pulled back over the cylinder $S^1\times \bbR^+$. When $d=3$, it is instead the space of sections of the homogeneous line bundle of degree $n$ which is pulled back over the cylinder $S^2\times \bbR^+$. The representations labeled $G^\pm_n$ and $G_{n,m_0}$ are the quantisations of the generic orbits $\mathcal{G}^\pm_{\lambda_0}$ and $\mathcal{G}_{n,\lambda_0}$ respectively. They are carried by the space of functions over the plane or half-plane. For all the indecomposable groups, the coadjoint orbits and UIRs are in exact correspondence. By considering the infinitesimal group action, one arrives at the equations governing the quantum system. For example, the group action for the representation $F^2_{n,E_0}$ of $A^z_3$, $d=3$ is
\begin{equation}
        [\sg(R,b,\boldsymbol{v},t)\cdot\psi](\zeta,q)= (\frac{\bar{a}-b\zeta}{|\bar{a}-b\zeta|})^{-n}e^{i\boldsymbol{p}(\zeta,q)\cdot\boldsymbol{v}}e^{iE_0 q^z t}\psi(\frac{a\zeta+\bar{b}}{-b\zeta+\bar{a}},e^{b}q)
\end{equation}
Defining
\begin{equation}
    \begin{split}
        \hat{H}\psi&:=-i\frac{d}{dt}\sg(1,0,\boldsymbol{0},t)\cdot \psi |_{t=0}\\
        \hat{P}_1 \psi&:=-i\frac{d}{dv_1}\sg(1,0,v_1,0,0)\cdot \psi|_{v_1=0}\\
        \hat{P}_2 \psi&:=-i\frac{d}{dv_2}\sg(1,0,0,v_2,0)\cdot \psi|_{v_2=0}\\
    \end{split}
\end{equation}
and
\begin{equation}
    \hat{\boldsymbol{P}^2}\psi=(\hat{P}_1^2+\hat{P}_2^2)\psi
\end{equation}
we arrive at the quantum version of equation \ref{disp}:
\begin{equation}
    \hat{H}^2\psi=E_0^2 (\hat{\boldsymbol{P}^2})^z \psi
\end{equation}
\begin{subtables}
\begin{table}[H]
  \centering
  \caption{Unitary Irreducible Representations}
  \label{tab6a}
  \rowcolors{2}{blue!10}{white}
\begin{tabular}{c|c|c|c|c} 
 Group & UIR&Characters & Carrier Space&
 Integration Measure\\
 \midrule
 $A_3^z$, $d=2$ &$P_{n,\delta_0}$&$n\in \bbZ$, $\delta_0 \in \bbR$&$\bbC$& Discrete \\ 
    &$V^\pm_n$&$n\in \bbZ$&$L^2(\bbR^\pm,\bbC)$& $\frac{dq}{q}$ \\ 
    &$F^\pm_{E_0}$&$E_0\in \bbR$&$L^2(S^1\times \bbR^+,E_\pm)$& $\frac{d\theta  dq}{q}$ \\ 
\midrule
 $A_3^z$, $d=3$  &$P_{n,\delta_0}$&$n\in \bbN_0$, $\delta_0\in \bbR$&$\bbC^{n+1}$ &Discrete\\
    &$V_n^\pm$&$n\in \bbN_0$ &$L^2(\bbR^\pm,\bbC)$ &$\frac{dq}{q}$\\
    &$F^2_{n,E_0}$&$n\in \bbZ$, $E_0\in \bbR$&$L^2(S^2\times \bbR^+, \mathcal{O}(-n))$ & $\frac{d\zeta  d\bar{\zeta} dq}{(1+|\zeta|^2)^2 q}$\\
\midrule
 $A_7$  &$P_{n,\delta_0}$&$n\in \bbZ$, $\delta_0\in \bbR$&$\bbC$& Discrete\\ 
    &$TSV^\pm$&&$L^2(S^1\times \bbR^+,E_\pm)$& $\frac{d\theta dq}{q}$\\ 
    &$G^\pm_n$&$n\in \bbZ$&$L^2(\bbR\times \bbR^\pm,\bbC)$& $\frac{dx dy}{y}$ \\ 
\midrule
 $C_1^1$&$P_{n,\delta_0,E_0}$ &$n\in \bbZ$, $\delta_0\in \bbR$, $E_0\in \bbR$&$\bbC$& Discrete\\
    &$TS^\pm_{\delta_0,p_0,E_0}$ &$\delta_0\in \bbR$, $p_0\in \bbR^+$, $E_0\in\bbR$&$L^2(S^1,E_\pm)$& $d\theta$\\
    &$G_{n,m_0}$ &$n\in \bbZ$, $m_0\in \bbR\backslash\{0\}$&$L^2(\bbR^2,\bbC)$& $dx dy$\\
\midrule
 $C_3$, $d=2$&$P_{n,\delta_0,E_0}$&$n\in \bbZ$, $\delta_0\in \bbR$, $E_0\in \bbR$&$\bbC$&Discrete\\
    &$V_{n,m_0}$&$n\in \bbZ$, $m_0\in \bbR\backslash\{0\}$&$L^2(\bbR,\bbC)$&$dq$\\
    &$Y^\pm_{E_0,m_0}$&$E_0\in \bbR$, $m_0 \in \bbR$&$L^2(S^1\times \bbR^+,E_\pm)$&$\frac{d\theta dq}{q}$\\
\midrule
$C_3$, $d=3$&$P_{n,\delta_0,E_0}$&$n\in \bbN_0$, $\delta_0\in \bbR$, $E_0\in \bbR$&$\bbC^{n+1}$&Discrete\\
    &$V_{n,m_0}$&$n\in \bbN_0$, $m_0\in \bbR\backslash\{0\}$&$L^2(\bbR,\bbC)$&$dq$\\
    &$Y^2_{n,E_0,m_0}$&$n\in \bbZ$, $E_0\in \bbR$, $m_0\in \bbR$&$L^2(S^2\times \bbR^+,\mathcal{O}(-n))$&$\frac{d\zeta  d\bar{\zeta} dq}{(1+|\zeta|^2)^2 q}$\\
\bottomrule
\end{tabular}
\caption*{The UIRs are in exact correspondence with the coadjoint orbits, as is expected, apart from the fact that the orbits $\mathcal{SP}$ and $\mathcal{P}$, and $\mathcal{SV}$ and $\mathcal{V}$, which have different geometry, produce the same representations, $P$ and $V$. When the radius of the associated $S^2$ goes to zero, this corresponds to the trivial representation ($n=0$) of $\SU(2)$ inside $A_3^z$, $d=3$ and $C_3$, $d=3$. The trivial and M\"obius bundles over $S^1$ are denoted respectively by $E_+$ and $E_-$, while $\mathcal{O}(-n)$ denotes the homogenous bundle of degree $n$ over $S^2\cong \mathbb{CP}^1$.}
\end{table}
\begin{table}[H]
  \centering
  \caption{Unitary Irreducible Representations}
  \label{tab6b}
  \rowcolors{2}{blue!10}{white}
\begin{tabular}{c|c|c} 
 Group & UIR& Group Action\\
 \midrule
 $A_3^z$, $d=2$ &$P_{n,\delta_0}$&$\sg(R,b,\bv,t)\cdot \psi=\nu_n(R)e^{i\delta_0 b}\psi$\\ 
    &$V^\pm_n$&$[\sg(R,b,\bv,t)\cdot \psi](q)=\nu_n(R)e^{iqt}\psi(e^{bz}q)$\\ 
    &$F^\pm_{E_0}$& $[\sg(R,b,\bv,t)\cdot \psi](\bp)=Q(R,\bp)e^{i\bp\cdot \bv}e^{i E_0 |\bp|^z t} \psi(e^b R^{-1}\bp)$\\ 
\midrule
 $A_3^z$, $d=3$  &$P_{n,\delta_0}$& $\sg(R,b,\bv,t)\cdot \psi = \rho_n(R)e^{i\delta_0 b}\psi$\\
    &$V_n^\pm$&$ [\sg(R,b,\bv,t)\cdot \psi](q)=\rho_n(R)e^{iqt}\psi(e^{bz}q)$\\
    &$F^2_{n,E_0}$&$ [\sg(R,b,\boldsymbol{v},t)\cdot\psi](\zeta,q)= (\frac{\bar{a}-b\zeta}{|\bar{a}-b\zeta|})^{-n}e^{i\boldsymbol{p}(\zeta,q)\cdot\boldsymbol{v}}e^{iE_0 q^z t}\psi(\frac{a\zeta+\bar{b}}{-b\zeta+\bar{a}},e^{b}q)$\\
\midrule
 $A_7$  &$P_{n,\delta_0}$&$\sg(R,b,\bv,t)\cdot \psi = \nu_n(R)e^{i\delta_0 b}\psi$\\ 
    &$TSV^\pm$&$[\sg(R,b,\bv,t)\cdot \psi](\bp)=Q(R,\bp)e^{i\bp\cdot \bv}\psi(e^b R^{-1}\bp)$\\ 
    &$G^\pm_n$&$ [\sg(R,b,\bv,t)\cdot \psi_\pm](x,y)=\nu_n(R)e^{iyt+i|y|^\frac{1}{2}xv_2}[S(R)\psi_\pm] (x\pm |y|^\frac{1}{2}v_1,e^{2b}y)$\\ 
\midrule
 $C_1^1$&$P_{n,\delta_0,E_0}$ &$\sg(R,b,\bv,t,z)\cdot \psi = \nu_n(R)e^{i\delta_0 b}e^{iE_0t}\psi$\\
    &$TS^\pm_{\delta_0,p_0,E_0}$ &$[\sg(R,b,\bv,t,z)\cdot \psi](\bp)=Q(R,\bp)e^{i\delta_0 b}e^{i\bp\cdot \bv}e^{iE_0 t}\psi(R^{-1}\bp)$\\
    &$G_{n,m_0}$ &$ [\sg(R,b,\bv,t,z)\cdot \psi](x,y)=\nu_n(R)e^{i(m_0 z+yt+xv_2)}[S(R)\psi](x+m_0 v_1,y+m_0 b)$\\
\midrule
 $C_3$, $d=2$&$P_{n,\delta_0,E_0}$&$\sg(R,b,\bv,t,z)\cdot \psi=\nu_n (R)e^{i\delta_0 b}e^{iE_0 t}\psi$\\
    &$V_{n,m_0}$&$ [\sg(R,b,\bv,t,z)\cdot \psi](q)=\nu_n(R)e^{im_0 z+ iqt}\psi(q+bm_0)$\\
    &$Y^\pm_{E_0,m_0}$&$[\sg(R,b,\bv,t,z)\cdot \psi](\bp)=Q(R,\bp)e^{i\bp\cdot \bv}e^{i(E_0+m_0\log|\bp|) t}e^{im_0 z}\psi(e^b R^{-1}\bp)$\\
\midrule
$C_3$, $d=3$&$P_{n,\lambda_0,E_0}$&$\sg(R,b,\bv,t,z)\cdot \psi=\rho_n(R)e^{i\delta_0 b}e^{iE_0 t}\psi$\\
    &$V_{n,m_0}$&$[\sg(R,b,\bv,t,z)\cdot\psi](q)=\rho_n(R)e^{im_0 z+ iqt}\psi(q+bm)$\\
    &$Y^2_{n,E_0,m_0}$&$[\sg(R,b,\boldsymbol{v},t)\psi](\zeta,q)= (\frac{\bar{a}-b\zeta}{|\bar{a}-b\zeta|})^{-n}e^{i\boldsymbol{p}(\zeta,q)\cdot\boldsymbol{v}}e^{i(E_0 +m_0\log|\bp(\zeta,q)|) t}e^{im_0 z}\psi(\frac{a\zeta+\bar{b}}{-b\zeta+\bar{a}},e^{b}q)$\\
\bottomrule
\end{tabular}
\caption*{$\nu_n$ is the degree $n$ representation of $\U(1)$ while $\rho_n$ is the $n+1$ dimensional representation of $\SU(2)$. $Q(R,\bp)$ accounts for the possible twist in the vector bundle over $S^1$ or $S^1\times \bbR$. For more information, see equation \ref{Twist}.}
\end{table}
\end{subtables}
\section{Conclusion}
The results of this paper classify all coadjoint orbits and unitary irreducible representations of the Lifshitz groups and their one dimensional central extensions. The Lifshitz spacetimes were classified in \cite{MR4577534}. The next step is to realise the coadjoint orbits as classical particle trajectories and the UIRs as quantum fields. The first is done by considering a particle action on $G/(K_\chi \cap T)$ such that the curves obtained from  varying this action project to single points in $\mathcal{O}_\chi=G/G_\chi$. The projection of the curves to the spacetime $M:=G/K$ form classical Lifshitz particle trajectories. Projecting to a different point in the same coadjoint orbit corresponds to acting with a symmetry of the Lifshitz system. An example of such is done for the Galilei groups in \cite{Figueroa-OFarrill:2024ocf}. Going from UIRs to quantum fields requires going from a representation of $G$ to sections of a vector bundle on $M=G/K$. The group $G$ fibres over $M$ as a principal $K-$bundle so we look for representations of $K$ to form an associated bundle. There is a method for going from the UIRs of $G$ to representations of $K$ via a sort of group theoretic Fourier transform. Dimensional considerations often impose constraints which can be interpreted as field equations for the quantum fields. A good example of this can once again be found in \cite{Figueroa-OFarrill:2024ocf} for the Galilei groups. It may be of interest to realise some of the Lifshitz UIRs as field theories. Another direction of research can be found in the so called Boost Extended Lifshitz Lie Algebras (BELLAs). These are generated by extending the kinematical Lie algebras discussed in \cite{primer} by a dilation. A paper on this is forthcoming.
\appendix
\section{Induction of the representation $TS^2_{n,p_0}$ of $\Euc(3)$ from $D_{n,p_0}$}\label{app1}
We explain in more detail the process of inducing the representation $TS^2_{n,p_0}$ from the inducing representation of $\U(1)\times \bbR^3$:
\begin{equation}
    D_{n,p_0}(R,\bv)=v_n(R)e^{i\bp\cdot \bv}
\end{equation}
We have the fibration $\Euc(3)\rightarrow S^2_{p_0}$ over the sphere with radius $p_0$ and we choose a section $\sigma$ of this fibration. Since the action of the $\bbR^3-$component of $\Euc(3)$ on $S^2$ is trivial, we can choose the section $\sigma$ to be as follows. Let $\bp\in S^2_{p_0}$ and let $\Sigma$ be a section of the Hopf bundle $\SU(2)\rightarrow S^2_{p_0}$ such that $\Sigma(\boldsymbol{p})p_0 \boldsymbol{e}_1 =\boldsymbol{p}$, where $\Sigma$ acts in the vector representation. Then we define $\sigma$ by
        \begin{equation}
            \sigma(\bp)=\sg(\Sigma(\boldsymbol{p}),\boldsymbol{0})
        \end{equation}
Recall that this section allows us to locally write down the representation induced by $D_{n,p_0}$ as
\begin{equation}
            [\sg\cdot\psi](\boldsymbol{p}) = D_{n,p_0}(\sh(\sg^{-1},\boldsymbol{p})^{-1})\psi(\sg^{-1}\cdot\boldsymbol{p})
        \end{equation}
where $\sh(\sg,\bp)$ is the compensating gauge transformation defined by
\begin{equation}
    \sg \sigma(\bp)=\sigma(\sg\cdot \bp)\sh(\sg,\bp)
\end{equation}
The section $\Sigma: S^2_{p_0}\rightarrow \SU(2)$ has its own compensating gauge transformation $H$, defined by
\begin{equation}
    R\Sigma(\bp)=\Sigma(R\bp)H(R,\bp)
\end{equation}
for $R\in \SU(2)$ and $\bp \in S^2_{p_0}$. It is calculated that 
\begin{equation}
    \sh(\sg(R,\bv),\bp)=\sg(H(R,\bp),\Sigma(\bp)^{-1}\bv)
\end{equation}
so that the induced representation of $\Euc(3)$ is
    \begin{equation}
            [\sg(R,\boldsymbol{v})\cdot\psi](\boldsymbol{p})=\nu_n(\sg(H(R^{-1},\boldsymbol{p})^{-1})e^{ip_0\vec{e}_3\cdot\Sigma(\boldsymbol{p})^{-1}\boldsymbol{v}}\psi(R^{-1}\boldsymbol{p})
        \end{equation}
 Noting that $\Sigma(\boldsymbol{p})\cdot p_0\boldsymbol{e}_3=\boldsymbol{p}$ can write the action more neatly:
    \begin{equation}
            [\sg(R,\boldsymbol{v})\cdot\psi](\boldsymbol{p})=\nu_n(\sg(H(R^{-1},\boldsymbol{p})^{-1})e^{i\bp\cdot\boldsymbol{v}}\psi(R^{-1}\boldsymbol{p})
        \end{equation}
 We now switch to homogeneous stereographic coordinates $\zeta$ on the sphere $S^2$. $\boldsymbol{p}$ is now the embedding $\boldsymbol{p}(\zeta): S^2\rightarrow \bbR^3$ whose image is the sphere of radius $p_0$. We can now define the section $\Sigma: S^2\rightarrow \SU(2)$ as follows:
    \begin{equation}
        \Sigma(\zeta)= \frac{1}{\sqrt{|\zeta|^2+1}}\begin{pmatrix}
            1&-\bar{\zeta}\\
            \zeta&1
        \end{pmatrix}
    \end{equation}
    Consider a general $\SU(2)$ element
    \begin{equation}
        R= \begin{pmatrix}
        a&b\\
        -\bar{b}&\bar{a}
    \end{pmatrix}
    \end{equation}
    with $|a|^2+|b|^2 =1$. This gives the compensating gauge transformation
    \begin{equation}
        H(R,\boldsymbol{p}(\zeta)) = \frac{1}{|a+b\zeta|}\begin{pmatrix}
            a+b\zeta&0\\
            0&\bar{a}+\bar{b}\bar{\zeta}
        \end{pmatrix}\in \U(1)
    \end{equation}
    Noting that rotations act as conformal transformations on homogeneous coordinates, we arrive at the representation in homogenous coordinates:
    \begin{equation}
                [\sg(R,\boldsymbol{v})\cdot\psi](\zeta)= (\frac{\bar{a}-b\zeta}{|\bar{a}-b\zeta|})^{-n}e^{i\boldsymbol{p}(\zeta)\cdot\boldsymbol{v}}\psi(\frac{a\zeta+\bar{b}}{-b\zeta+\bar{a}})
     \end{equation}

     The invariant volume form that we use to define the inner product is the standard volume form induced by the standard metric on $S^2$, as this is rotationally invariant and the representation is thus unitary. In stereographic coordinates it is given by equation \ref{vol}.

\section{Determination of the intertwiner $S(R)$ for the Oscillator Group}\label{app2}
The action of $R \in \U(1)$ on a representation $\chi$ in $\hat{\N}$, which lies on the punctured vertical with non-zero $m_0$, is as follows. The representation
 \begin{equation}
    [\chi(\sg(v_1,v_2,\gamma))\cdot \psi](q)=e^{i(m_0\gamma+qv_2)}\psi(q+m_0v_1)
    \end{equation}
gets sent to
 \begin{multline}
    [\chi'(\sg(v_1,v_2,\gamma))\cdot\psi](q) = e^{im_0(\gamma-N^2v_1v_2+MN\frac{v_2^2-v_1^2}{2})}\\
    \times e^{iq(-Nv_1+Mv_2)} \psi(q+m_0(Mv_1+Nv_2)).
\end{multline}
where 
\begin{equation}
    R\bv=\begin{pmatrix}
        M & -N\\
        N & M
    \end{pmatrix}\bv
\end{equation}
The central element $\gamma$ still acts according to the character $m_0$ so these representations must be equivalent. We therefore seek a unitary intertwiner $S(R)$ such that $S\chi'=\chi S$. By Schur's lemma, if such an intertwiner exists, it must be unique up to a scalar multiple. We conjecture that
\begin{equation}
    [S(R)\psi](q)=\int_\bbR K(R,q,k)\psi(k)dk
\end{equation}
for some integration kernel $K$. The intertwiner must also intertwine with the algebra action. To that end we define
\begin{equation}
    \begin{split}
        [\hat{Q}\psi](q) &:= \frac{d}{dv_1}[\chi(\sg(v_1,0,0))\psi](q)|_{v_1=0}\\
[\hat{P}\psi](q) &:= \frac{d}{dv_2}[\chi(\sg(0,v_2,0))\psi](q)|_{v_2=0}\\
[\hat{Q}'\psi](q) &:= \frac{d}{dv_1}[\chi'(\sg(v_1,0,0))\psi](q)|_{v_1=0}\\
[\hat{P}'\psi](q) &:= \frac{d}{dv_2}[\chi'(\sg(0,v_2,0))]\psi(q)|_{v_2=0}\\
    \end{split}
\end{equation}
The requirements that 
\begin{equation}
\begin{split}
    S\hat{Q}'&=\hat{Q}S\\
    S\hat{P}'&=\hat{P}S
\end{split}
\end{equation}
lead to the equations
\begin{equation}
\begin{split}
    -ikNK-m_0 M\frac{dK}{dk}&=m_0\frac{dK}{dq}\\
    ikMK-m_0 N\frac{dK}{dk}&=iqK
\end{split}
\end{equation}
which have the solution
\begin{equation}
    K(R,q,k)=A(R)e^{\frac{iM}{2Nm_0}(q^2+k^2)-\frac{iqk}{Nm_0}}
\end{equation}
for a function $A(R)$ to be determined. The function is fixed by requiring $S(1)=1$, namely that
\begin{equation}
    \lim_{N\rightarrow 0}\int_{\bbR}K(R,q,k)\psi(k)dk=\psi(q)
\end{equation}
To determine the limit, we make a change of variables $u=\frac{k}{\sqrt{|N|}}-\frac{q}{M\sqrt{|N|}}$ so that
\begin{equation}
    \int_\bbR K(R,q,k)\psi(k)dk=A(R)\sqrt{|N|}e^{-\frac{iNq^2}{2Mm_0}}\int_\bbR e^{\sgn{N}\frac{iM}{2m_0}u^2}\psi(\sqrt{|N|}u+\frac{q}{M})du
\end{equation}
Let $C(R)=A(R)\sqrt{|N|}$.
Then
\begin{equation}
    \lim_{N\rightarrow 0} \int_\bbR K(R,q,k)\psi(k)dk=\lim_{N\rightarrow 0}C(R) \int_\bbR e^{\sgn{N}\frac{i}{2m_0}u^2}\psi(q)du
\end{equation}
The integral on the right is a simple Gaussian so we can solve for the form of $C$ so that the limit goes to $\psi(q)$. We arrive at the required expression for $A(R)$:
\begin{equation}
    A(R)=\frac{e^{-i\sgn{N}\sgn(m_0)\frac{\pi}{4}}}{\sqrt{2\pi |m_0||N|}}
\end{equation}
The kernel $K(R,q,k)$ is known as the Mehler kernel and it is the heat kernel for the Schr\"odinger equation. 

\section*{Acknowledgments}
The author would like to thank the EPSRC for funding during his PhD as
well as his supervisor, José Figueroa-O'Farrill, for suggesting the
problem and for useful discussions.


\providecommand{\href}[2]{#2}\begingroup\raggedright\endgroup

\end{document}